\documentclass{jfm}
\usepackage{graphicx}
\usepackage{epstopdf, epsfig}
\usepackage{hyperref}
\usepackage[usenames,dvipsnames,svgnames,table]{xcolor}
\usepackage{libertine}
\usepackage{mathptmx} 
\usepackage[utf8]{inputenc}
\usepackage{newtxmath}
\usepackage{pgfplots}
\usepackage{natbib}
\usepackage{microtype}
 
\usetikzlibrary{external}
\hypersetup{
colorlinks=true,
linkcolor=blue,
urlcolor=blue,
filecolor=blue,
citecolor=blue
}

\makeatletter

\patchcmd{\NAT@citex}
  {\@citea\NAT@hyper@{%
     \NAT@nmfmt{\NAT@nm}%
     \hyper@natlinkbreak{\NAT@aysep\NAT@spacechar}{\@citeb\@extra@b@citeb}%
     \NAT@date}}
  {\@citea\NAT@nmfmt{\NAT@nm}%
   \NAT@aysep\NAT@spacechar\NAT@hyper@{\NAT@date}}{}{}

\patchcmd{\NAT@citex}
  {\@citea\NAT@hyper@{%
     \NAT@nmfmt{\NAT@nm}%
     \hyper@natlinkbreak{\NAT@spacechar\NAT@@open\if*#1*\else#1\NAT@spacechar\fi}%
       {\@citeb\@extra@b@citeb}%
     \NAT@date}}
  {\@citea\NAT@nmfmt{\NAT@nm}%
   \NAT@spacechar\NAT@@open\if*#1*\else#1\NAT@spacechar\fi\NAT@hyper@{\NAT@date}}
  {}{}

\makeatother

\defcitealias{companion}{GMS}

\newcommand\ie{i.e.\ }

\newcommand{\vect}[1]{\boldsymbol{#1}}
\newcommand{\matr}[1]{\mathsfbi{#1}}

\newcommand\Fr{\mbox{\textit{Fr}}} 
\newcommand\im{\mathrm{i}\mkern1mu} 

\title{Linear stability of shallow morphodynamic flows}
\shorttitle{Linear stability of shallow morphodynamic flows}

\shortauthor{J. Langham, M. J. Woodhouse, A. J. Hogg, J. C. Phillips}
\author{Jake Langham\aff{1,2}
  \corresp{\email{j.langham@bristol.ac.uk}},
  Mark J. Woodhouse\aff{2},
  Andrew J. Hogg\aff{1}
 \and Jeremy C. Phillips\aff{2}}

\affiliation{\aff{1}School of Mathematics, Fry Building, University of Bristol,
Bristol, BS8 1UG, UK
\aff{2}{School of Earth Sciences, Wills Memorial Building, University of Bristol, Bristol, BS8 1RJ, UK}
}

\begin{document}

\maketitle

\begin{abstract}
It is increasingly common for models of shallow-layer overland flows to include
equations for the evolution of the underlying bed (morphodynamics) and the
motion of an
associated sedimentary phase. We investigate the linear stability properties of
these systems in considerable generality.  Naive formulations of the
morphodynamics, featuring exchange of sediment between a well-mixed
suspended load and the bed, lead to mathematically ill-posed governing
equations. This is traced to a singularity in the linearised system at Froude
number $\Fr = 1$ that causes unbounded unstable growth of short-wavelength
disturbances. The inclusion of neglected physical processes can restore well
posedness. Turbulent momentum diffusion (eddy viscosity) and a suitably
parametrised bed load sediment transport are shown separately
to be sufficient in this regard.  However, we demonstrate that such models
typically inherit an associated instability that is absent from
non-morphodynamic settings.  Implications of our analyses are considered for
simple generic closures, including a drag law that switches between fluid and
granular behaviour, depending on the sediment concentration. Steady
morphodynamic flows bifurcate into two states: dilute flows, which are stable at
low $\Fr$, and concentrated flows which are always unstable to disturbances in
concentration. By computing the growth rates of linear modes across a wide
region of parameter space, we examine in detail the effects of specific model
parameters including the choices of sediment erodibility, eddy viscosity and bed
load flux. These analyses may be used to inform the ongoing development of
operational models in engineering and geosciences.
\end{abstract}

\section{Introduction}
\label{sec:intro}
The growth of instabilities of inclined overland flows can cause small variations in
the free surface to roll up into large-amplitude waves and
shocks~\citep{Dressler1949,Needham1984}, with the potential over long distances
to turn a homogeneous flowing layer into a sequence of destructive
surges~\citep{Zanuttigh2007}.
These \emph{roll waves} have been observed to develop in shallow flows with
diverse rheologies, including turbulent fluid layers
\citep{Cornish1934,Needham1984,Balmforth2004}, 
hyperconcentrated suspensions and debris flows
\citep{Pierson1985a,Davies1986,Davies1992},
dense granular flows
\citep{Forterre2003,Razis2014} 
and mixtures of cohesive sediment \citep{Coussot1994,Ng1994}.
The appearance (or lack) of roll waves on volcanic debris flows (lahars) and
their waveform characteristics have been used to infer flow properties and
initiation processes \citep[e.g.][]{Doyle2010}.
When flows are able to erode and deposit material, additional modes of instability may be present, caused
by coupling between the flow and its underlying topography.  These
interactions, usually referred to as \emph{morphodynamics}, bring about a rich
collection of intriguing wavy bed patterns, formed in different physical
regimes \citep{Engelund1982,Seminara2010,Slootman2020}.
Where flows constitute dangerous natural hazards, morphodynamic uptake of mass
may significantly amplify their destructive power and therefore cannot be
ignored in geophysical models of these systems \citep{Iverson2015}. Post-event
structures in deposits have been interpreted as preservation of instabilities
during such flows \citep{Baloga2005}.

There has been considerable interest in mathematical stability problems thought
to underpin and give rise to these various phenomena.  The simplest relevant
setting is one-dimensional uniform shallow layers of turbulent water, flowing
down a constant incline.  Linear stability of these states depends on a single
control parameter, the Froude number, defined by $\Fr = \tilde u_0 / (g_\perp
\tilde h_0)^{1/2}$, where $\tilde h_0$, $\tilde u_0$ are the height and velocity
of the steady uniform flow, and $g_\perp$ denotes gravitational acceleration
resolved perpendicular to the slope.  For example, when the typical Ch\'ezy
formula for basal drag applies, the flow is unstable for all $\Fr >
2$~\citep{Jeffreys1925}.  Similar problems have been tackled over the years,
using different flow models and approaches to investigate various physical
systems.  The literature concerning the linear stability of such flows is vast.
It is particularly worth noting the breadth of settings that may be treated by
considering the evolution of small disturbances in the shallow-flow equations,
which includes turbulent open
water~\citep{Keulegan1940,Craya1952,Dressler1953,Thual2010}, mudflows on
impermeable \citep{Ng1994,Liu1994} and porous slopes \citep{Pascal2006}, debris
flows \citep{Zanuttigh2004} and granular flows
\citep{Forterre2003,GrayEdwards2014}.

The inclusion of morphodynamic processes adds complexity, but has nevertheless
received considerable attention, since stability theory provides a natural way
to investigate the genesis of observed bed patterns and surface waves. In this
case, the shallow-flow equations are paired with an equation for the bed
evolution and an appropriate description of how the flow and bed are coupled.
Depending on the application, different degrees of sophistication are needed. In
many contexts, the bed evolves slowly (relative to the flow velocity) and the
pattern-forming instabilities of its free surface may be explained
using analyses that assume a steady flow~\citep{Richards1980,Engelund1982}.
Where there is significant exchange of material over flow time scales, such as
in powerful debris flows~\citep{Hungr2005}, a fuller analysis is required, as
there is a strong two-way coupling between the flow and bed motion.

\cite{Trowbridge1987} identified the value of taking a generalised approach to
shallow-flow stability analysis, deriving a simple linear stability
criterion for any inclined uniform solution to the unidimensional shallow-flow
equations in the non-erosive case, subject to an arbitrary basal drag law.  In
doing so, the linear response of many different model rheologies was
encompassed.  This analysis was recently extended by \cite{Zayko2019}, who
showed that for some rheologies, Trowbridge's stability criterion is bypassed by
oblique (\ie non-slope-aligned) disturbances.  For morphodynamic flows, it seems
doubtful that comparably simple stability criteria may be obtained, due to the
presence of extra modes associated with the bed dynamics that complicate the
general picture.  However, operational models feature many different physical
closures for the various morphodynamic processes and in each case there is a
proliferation of viable choices.  Therefore, 
in this paper we formulate our analysis in a general setting so that our results
may then be applied to a 
variety of individual models.
We pay particular attention to a popular class of models
recently developed to describe events that feature rapid and
substantial transfer of material with the bed, such as violent dam breaks or
natural debris flows. This is achieved by augmenting the standard shallow-flow
equations with a transport equation for a `suspended load' of entrained solids
and a bed evolution equation featuring erosion and deposition
terms~\citep[e.g.][]{Cao2004,Cao2017}. The extent to which the sediment dynamics
affects stability of flows in this setting is not well understood.  Therefore,
we spend the bulk of this study attempting to address this in a general way. 

Stability analysis can reveal underlying shortcomings in a model.
In river morphodynamics, it is common practice to couple the Saint-Venant
equations with one or more `bed load' transport equations to describe the
dynamics of different sediment layers. It is now known that this
approach can lead to systems of non-hyperbolic governing equations that are ill
posed as initial value
problems~\citep{Cordier2011,Stecca2014,Chavarrias2018,Chavarrias2019}. Where
this occurs, these models are rendered inappropriate as descriptions of
dynamical flows, at least in the form typically used in numerical solvers.
Likewise, we shall prove that models with suspended sediment load are, in their
most basic formulation, ill posed when the Froude number is unity.
Two physical processes: turbulent diffusivity and bed load transport, are shown
separately to remove ill posedness. The former does so unconditionally; for the
latter, we derive general constraints for well-posed models similar to prior
analyses undertaken in the fluvial
setting~\citep{Cordier2011,Stecca2014,Chavarrias2018}.  By investigating the
posedness and stability of these extended formulations in a general setting,
with both bed and suspended load, we take steps towards a unified understanding
of shallow morphodynamic models across multiple flow regimes.
Moreover, it should be straightforward to apply our conclusions to individual
models, or to incorporate additional modelling terms into the analysis.


\section{Formulation}
\label{sec:formulation}
\begin{figure}
 \centering%
 \includegraphics{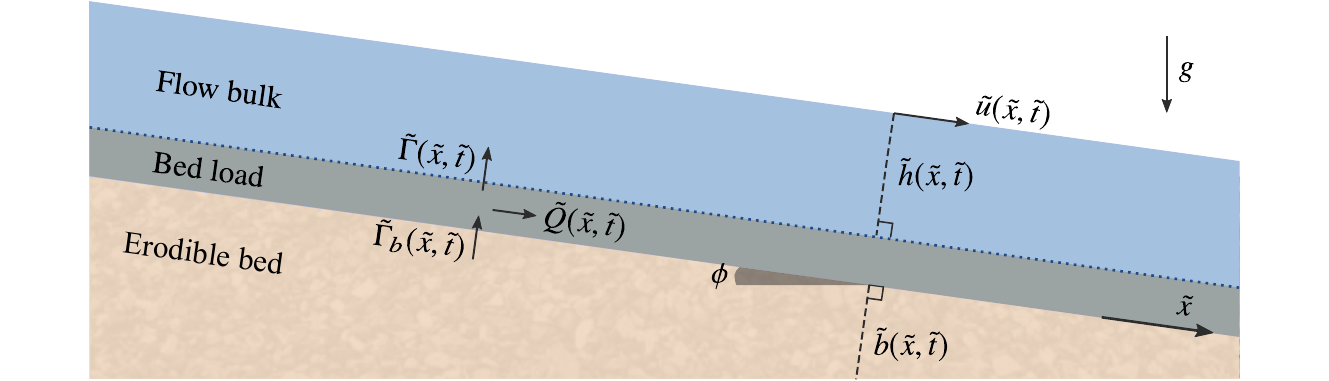}%
 \caption{%
     Diagram of the system under consideration. A shallow flowing layer of
     height~$\tilde h$ and velocity~$\tilde u$ travels along an initially
     uniform slope inclined at an angle~$\phi$ to the horizontal.  
     Underneath is a bed of height~$\tilde{b}$, composed of homogeneous sediment
     that may be carried as a distinguished load (of fixed depth) along the bed
     surface, or entrained into the flow bulk. The material transfer variables
     are labelled with arrows, to indicate the directions of positive transport.
 }%
 \label{fig:setup}
\end{figure}

The setting for this paper is the geometry depicted in figure~\ref{fig:setup},
which shows a cross-section of a free-surface flow at time $\tilde t$,
travelling down a sloping erodible bed principally driven by gravitational
acceleration $g$.
We fix a coordinate $\tilde x$, oriented along the slope, which is inclined at
a constant angle~$\phi$ to the horizontal.  Only motions and spatial variations
in the flow fields along this axis are considered. Both the flow height $\tilde
h(\tilde{x}, \tilde t)$ and bed height $\tilde b(\tilde x, \tilde t)$ are
measured in the direction normal to the slope and the depth of flowing material
is everywhere
assumed to be small, relative to its streamwise and lateral coverage 
along the slope plane. 

Governing equations for flows in this setting may be obtained by integrating the
continuity and momentum transport equations for a general continuum body over
the flow depth and
neglecting terms that are small for a shallow layer. This standard procedure
eliminates both the slope-normal components of motion and any non-hydrostatic
pressure gradients, and replaces the downslope velocity with its depth-averaged
value, denoted herein by $\tilde u(\tilde x, \tilde t)$.  Allowing for
linear-order variations in the bed gradient results in a contribution to the
depth-averaged hydrostatic pressure term only.  Higher-order variations (\ie
curvatures) may be considered, but these are not relevant for studying the
linear stability of flows on constant slopes.  For simplicity, we also choose to
omit `shape factors' -- free parameters arising from the depth integration that
quantify the level of vertical shear in the velocity profile. While these can,
in certain cases, modify solutions significantly~\citep{Hogg2004}, they are
typically unknown and very often neglected in modelling studies \citep[for
example]{Macedonio1992,Iverson1997,Cao2004,Xia2010}. Nevertheless, our analysis
could in principle be adapted to include them. 

If there are no morphodynamic processes present, the depth-averaged flow
density~$\tilde \rho(\tilde x, \tilde t)$ is a constant
field and the equations of motion are:
%
\begin{subequations}
\begin{gather}
    \frac{\partial \tilde h}{\partial \tilde t} +
    \frac{\partial ~}{\partial \tilde x}\left(\tilde h \tilde u\right)
    = 0,\label{eq:sw mass}\\
    \frac{\partial ~}{\partial \tilde t}\left( \tilde h\tilde u\right)
    +
    \frac{\partial ~}{\partial \tilde x}\left( \tilde h \tilde u^2 \right)
    + g \tilde h \cos \phi \frac{\partial ~}{\partial \tilde x}  \left(\tilde h
    + \tilde b \right)
    =
     g \tilde h \sin \phi -\tilde \tau / \tilde \rho.\label{eq:sw mom}
\end{gather}
\label{eq:sw all}%
\end{subequations}
%
The final component of~\eqref{eq:sw mom} is a forcing term obtained from
depth integration of the material stresses. It is a free constitutive law that
captures the aggregate rheology of the flow. 
%
A typical example is to set $\tilde \tau \propto \tilde u^2$, which models
the turbulent drag experienced by a fluid moving over a rough surface, although
there are many other choices.
To encompass a broad range of systems in our analysis, we take
$\tilde\tau$ to be an arbitrary function of the local flow fields.

We now allow the flow to exchange fluids and solids with the underlying bed, whose
height $\tilde b(\tilde{x}, t)$ is measured in line with $\tilde h$.  Entrained
solid material is assumed to be composed of homogeneous particles of
density~$\tilde \rho_s$ that are much smaller than the
flow depth, so that they may be treated as a continuous phase occupying a
(depth-averaged) fraction $\tilde\psi(\tilde{x}, \tilde t)$ of the flow volume.
The remainder of the mixture (occupying fraction $1-\tilde \psi$) is fluid of
constant density~$\tilde \rho_f$.  The overall density of the flow
is then
\begin{equation} 
    \tilde\rho = \tilde \rho_f (1 - \tilde\psi) + \tilde \rho_s \tilde\psi.
    \label{eq:density}
\end{equation}

The volumetric flux of net mass (comprising both fluid and solid phases)
transferred to the flow bulk from below shall be denoted
by~$\tilde\Gamma(\tilde{x}, \tilde t)$.  This function encapsulates the
competing processes of sediment entrainment and deposition into a single source
term. (Example parametrisations of these processes are given later,
in~\S\ref{sec:closures}.) When $\tilde \Gamma > 0$, there is net uptake of
material into the suspended load of the bulk; when $\tilde \Gamma < 0$, there is
a net loss.  On including the contribution of this term equation~\eqref{eq:sw
mass}, which describes conservation of the total flow mass, becomes
\begin{subequations}
\begin{gather}
    \frac{\partial \tilde h}{\partial \tilde t} 
    + \frac{\partial~}{\partial \tilde x}(\tilde h \tilde u)
    =
    \tilde \Gamma.\label{eq:sw ero mass}
\end{gather}%

We assume the bed has constant density~$\tilde \rho_b$ and is
everywhere saturated, comprising a homogeneous mixture of fluid and solids, with
the latter phase occupying volumetric fraction~$\tilde \psi_b$.
The volumetric flux of the
solid and fluid phases into the flow bulk are then necessarily~$\tilde\psi_b\tilde\Gamma$ and~$(1-\tilde \psi_b)\tilde\Gamma$ respectively.
This leads to a separate mass conservation equation for the solid phase:
\begin{gather}
    \frac{\partial ~}{\partial \tilde t}(\tilde\psi \tilde h) 
    + \frac{\partial~}{\partial \tilde x}(\tilde \psi \tilde h \tilde u)
    =
    \tilde\psi_b \tilde \Gamma.\label{eq:sw ero solid mass}
\end{gather}%
Between the flowing layer and the bed, we allow for a distinguished mobile layer
of material, commonly referred to as the \emph{bed load}, that travels with flux
$\tilde Q(\tilde x, \tilde t)$. Below this layer, the underlying substrate is
assumed to be immobile and transfers material to the bed load at a rate $\tilde
\Gamma_b$, such that the bed height obeys $\partial \tilde b / \partial \tilde t
= -\tilde \Gamma_b$.  If the middle bed load layer possesses a constant
characteristic thickness, its mass conservation relation is given by simply
$\partial \tilde Q / \partial \tilde x = \tilde \Gamma_b - \tilde \Gamma$.
(Figure~\ref{fig:setup} is a useful reference for the sign conventions of the
fluxes and source terms here.) Therefore, conservation of mass for the moving
and immobile components of the bed as a whole implies
\begin{equation}
    \frac{\partial \tilde b}{\partial \tilde t}
    + \frac{\partial \tilde Q}{\partial \tilde x}
    =
    -\tilde\Gamma.
    \label{eq:mass transfer}%
\end{equation}
%
The inclusion of bed load conceptually separates the gradual crawl of grains
along the bed surface (as typically observed in fluvial systems, for example),
from transfer of sediment with the bulk flow. The latter process, through
changes to the bulk density and drag characteristics, affects the dynamics of the
overlying flow. Since these processes are commonly modelled by flux and source
terms respectively, they cannot be combined in our analysis.

To complete the morphodynamic description, the momentum conservation
equation~\eqref{eq:sw mom} must be amended to account for spatial variations in
$\tilde \rho$ that may arise via the transport dynamics of the solids fraction.
Re-deriving~\eqref{eq:sw mom} from the morphodynamic standpoint introduces a
density dependence into each term and also leads to an extra contribution
$\tilde \rho \tilde u_b \tilde \Gamma$, included in some models, that
accounts for jumps in velocity, stress and density between the flow and the
layer beneath it, which necessarily occur when particles are either mobilised or
de-entrained.  In the absence of bed load, this term represents the rate of
change of momentum required to accelerate the entrained material to a
characteristic slip velocity~$\tilde u_b(\tilde h, \tilde u, \tilde \psi)$ near
the bed surface. A comprehensive derivation and discussion of this term is given
by~\cite{Iverson2015}.
The complete governing equation for momentum can be written as
\begin{equation}
    \frac{\partial~}{\partial \tilde t}(\tilde \rho \tilde h \tilde{u}) 
    + \frac{\partial~}{\partial \tilde x}\left(
    \tilde \rho \tilde h \tilde u^2 \right)
    + \frac{1}{2} g \cos \phi \frac{\partial~}{\partial \tilde x}\left( \tilde \rho \tilde h^2
    \right)
    =
    \tilde \rho g \tilde h \left(\sin \phi 
    - \cos \phi \frac{\partial \tilde b}{\partial \tilde x}
    \right)
    - \tilde\tau
    +\tilde \rho \tilde{u}_b \tilde\Gamma.\label{eq:sw ero mom}
\end{equation}\label{eq:sw morpho}%
\end{subequations}

Equations~(\ref{eq:sw morpho}\emph{a--d}) constitute a general shallow-water
model for a sediment-carrying flow, coupled with its underlying topography by
closures for mass exchange and bed load flux. 
Our goal is to understand some of the general properties of these models, the
solutions of the governing equations and their stability. 
We divide this overall framework into four subcategories:
\begin{enumerate}
    \item \emph{Hydraulic limit.} 
        When $\tilde \Gamma = \tilde Q = 0$,~(\ref{eq:sw morpho}\emph{a--d})
        reduce to equations~(\ref{eq:sw all}\emph{a,b}), which are appropriate
        for flows on inerodible substrates.
        These have been thoroughly studied elsewhere and provide a
        useful reference point for the other cases.
        We briefly cover their linear stability 
        in~\S\ref{sec:hydraulic limit}.
    \item \emph{Suspended load model.}
        When $\tilde \Gamma \neq 0$ and $\tilde Q = 0$, any eroded sediment is
        entrained directly into the bulk flow. This is our primary focus in the
        paper. Models in this class are employed to describe energetic flows
        with significant sediment uptake and mixing, often leading to high
        solids concentrations.
        Recent example studies from the literature 
        include (but are not limited to) \cite{Cao2004,Cao2006,Wu2007,Yue2008}
        and~\cite{Li2011}.
        We derive general linear stability results for these models
        in~\S\ref{sec:bed exchange} and~\S\ref{sec:regularisation}; existence of
        steady solutions and their stability properties are explored in detail
        for an example model in~\S\ref{sec:existence}--\S\ref{sec:effect of
        regularisation}.
    \item \emph{Bed load model.}
        When $\tilde \Gamma = 0$ and $\tilde Q \neq 0$, eroded sediment is
        only carried in the distinguished bed load layer.  These models are most
        often used in fluvial settings, where the effects of lateral sediment
        transport are important, but individual grains receive little upward
        momentum and remain largely near the bed surface. 
        These models are widely used: a partial list of examples in the
        literature includes \cite{Hudson2005, Murillo2010, Benkhaldoun2011,
        Siviglia2013, Juez2014, Kozyrakis2016}. 
    \item \emph{Combined model.}
        A few recent studies allow for both $\tilde \Gamma \neq 0$ and $\tilde Q
        \neq 0$, including \cite{Wu2007,Liu2015,Liu2017} and a two-layer model
        due to~\cite{Swartenbroekx2013} (which includes a momentum equation for
        the bed load layer and is therefore not strictly encompassed herein).
        This is approach is less commonplace, but potentially useful for
        physical situations that fall between the regimes of~(ii) and~(iii).
        Moreover, as we suggest below, it may be more widely applicable as a way
        to address issues with the formulation of suspended load models.  We
        analyse the well posedness of these models together with pure bed load
        models in~\S\ref{sec:bed load}. Existence of steady states for example
        closures in the combined model is analysed in~\S\ref{sec:existence} and
        their linear stability is explored in~\S\ref{sec:effect of bed load}.
\end{enumerate}

While very many models fit our general framework, there are a few
underlying assumptions that are important to list, since they dictate the scope
of our analysis.
We have already made explicit
our requirement that the flow and bed are composed of small, roughly homogeneous
grains, so that the solid fraction may be treated as a single continuous phase.
Moreover, we have neglected the equations for bed load momentum (usually
considered negligible) and the solid phase momentum, which may be combined with
that of the overall mixture provided the flow is well mixed.
Amongst other physical effects, we have implicitly neglected the role of
interstitial pore fluid pressures between grains, whose dynamics couples with
shear and dilation of the granular phase~\citep{Guazzelli2018}. These
interacting processes can lead to dramatic transients known to impact flow
outcomes and cause debris flows to be sensitive to initiation
conditions~\citep{Iverson1997,Iverson2000}. Consequently, our analysis is only
strictly relevant to flow regimes where pore pressure is negligible (\ie less
concentrated flows), or situations where the system has everywhere relaxed to
the ambient hydrostatic pressure. 

\section{Linear stability}
\label{sec:erosive}
We assume the presence of a uniform steady flowing layer of height $\tilde h_0$,
velocity $\tilde u_0$, solid fraction $\tilde\psi_0$, density $\tilde \rho_0 =
\tilde \rho(\tilde\psi_0)$, travelling on a flat sloping bed of (arbitrary)
height $\tilde b_0$. According to~(\ref{eq:sw morpho}\emph{a--d}), the existence of
such a solution depends on the particular parametrisations for drag and solids
exchange, which must satisfy
\begin{subequations}
\begin{gather}
    \tilde \tau(\tilde h_0, \tilde u_0, \tilde\psi_0) = 
    \tilde \rho_0 g \tilde
    h_0 \sin\phi
    ~~\mathrm{and}~~
    \tilde \Gamma(\tilde h_0, \tilde u_0,
    \tilde\psi_0) = 0.
    \tag{\theequation\emph{a,b}}%
\end{gather}
\label{eq:steady morpho balance}%
\end{subequations}
That is, at steady state, gravitational forcing is exactly balanced by the basal
drag and there is no net mass transfer between the bed and the flow.  We may
linearise the governing equations around these putative steady flows without
making explicit choices for $\tilde \tau$ and $\tilde \Gamma$. The bed load
$\tilde Q$ may also be kept as a general unknown function. In doing so, we
obtain general expressions that can be adapted to different
situations by inputting appropriate closures.  Detailed discussion
of the existence of steady flows, specialised to the case of fluid--grain
mixtures, is given later, in~\S\ref{sec:existence}.

For simplicity, we choose to rescale length, time and the dynamical variables as
\begin{subequations}
\begin{gather}
x = \tilde{x} /
\tilde{\ell}_0,~
t = \tilde t \tilde{u}_0 / \tilde{\ell}_0,~
h = \tilde{h} / \tilde{h}_0,~
u = \tilde{u} / \tilde u_0,~
\psi = \tilde \psi / \tilde \psi_b~\text{and}~
b = \tilde b / \tilde{h}_{0},
\tag{\theequation\emph{a--f}}%
\end{gather}
where $\tilde\ell_0 \equiv \tilde{u}_{0}^2 / (g\sin\phi)$.
Additionally, we define
\begin{gather}
\tau = \tilde \tau/\tilde{\tau}_0,~
\Gamma = \tilde \Gamma \tilde{\ell}_0/(\tilde h_0 \tilde u_0),~
Q = \tilde Q / (\tilde h_0 \tilde u_0),~
u_b = \tilde u_b / \tilde u_0,~
\rho = \tilde \rho / \tilde \rho_0,~
\mathrm{and}~
\rho_i = \tilde\rho_i / \tilde \rho_0,
    \tag{\theequation\emph{g--l}}%
\end{gather}
\label{eq:variable subs all}%
\end{subequations}
for $\tilde\rho_i \in \{ \tilde\rho_b, \tilde\rho_f, \tilde\rho_s\}$
and $\tilde\tau_0 \equiv \tilde\rho_0 g\tilde{h}_0 \sin\phi$.
On substituting~(\ref{eq:variable subs all}\emph{a--l}) into the governing
equations~(\ref{eq:sw morpho}\emph{a--d}) and simplifying, one arrives at
\begin{subequations}
\begin{align}
    \frac{\partial h}{\partial t}
    + \frac{\partial~}{\partial x}(hu)
    &=
    \Gamma,
    \label{eq:sw morpho nondim 1}\\
    \frac{\partial~}{\partial t}(\psi h)
    + \frac{\partial~}{\partial x}(\psi hu)
    &=
    \Gamma,\label{eq:sw morpho nondim 2}\\
    \frac{\partial~}{\partial t}(\rho h u)
    + \frac{\partial~}{\partial x}\left(\rho h u^2
    + \frac{1}{2}\Fr^{-2} \rho h^2 \right)
    &=
    \rho h \left(1 -\Fr^{-2} \frac{\partial b}{\partial x} \right)
    - \tau
    + \rho u_b \Gamma,
    \label{eq:sw morpho nondim 3}\\
    \frac{\partial b}{\partial t} 
    + \frac{\partial Q}{\partial x} &= -\Gamma,
    \label{eq:sw morpho nondim 4}
\end{align}
\label{eq:sw morpho nondim}%
\end{subequations}
where $\Fr \equiv \tilde u_0 / (g\tilde h_0 \cos\phi)^{1/2}$ is the Froude number of
the steady flow.

In this rescaled problem, the steady flow is a solution of~\eqref{eq:sw
morpho nondim 1}--\eqref{eq:sw morpho nondim 4} with height $h_0 = 1$, velocity
$u_0 = 1$, solid fraction $\psi_0 = \tilde \psi_0 / \tilde \psi_b$ and arbitrary
bed height $b_0$. The density of the layer is $\rho_0 = 1$.
%
Any slope-aligned perturbation to this state may be decomposed into individual
Fourier modes of real wavenumber $k$, which grow or decay in time at some
unknown complex growth rate~$\sigma$.
To find a general formula for $\sigma$, we construct the following ansatz:
\begin{subequations}
    \begin{gather}
        h(x, t) = 1 + \epsilon h_1 \exp(\sigma t + \im kx),\label{eq:ansatz 1}\\
    u(x, t) = 1 +
\epsilon u_1 \exp(\sigma t + \im kx),\\
    \psi(x, t) = \psi_0 + \epsilon \psi_1
\exp(\sigma t + \im kx), \\
    b(x, t) = b_0 + \epsilon b_1 \exp(\sigma t + \im
        kx), \label{eq:ansatz 4}
    \end{gather}
\end{subequations} 
where $h_1, u_1, \psi_1, b_1$ are unknown constants and $\epsilon \ll 1$.
By substituting~\eqref{eq:ansatz 1}--\eqref{eq:ansatz 4} into~(\ref{eq:sw morpho
nondim}\emph{a--d}) and dropping $O(\epsilon^2)$ terms, we obtain a
linear system of the form 
\begin{equation}
    \sigma\matr{A}\vect{q} + \im k \matr{B}\vect{q} + \matr{C}\vect{q} =
    \vect{0}, 
    \label{eq:morpho eigenproblem}
\end{equation}
where $\vect{q} = (h_1, u_1, \psi_1, b_1)^T$, and $\matr{A}$, $\matr{B}$,
$\matr{C}$ are $4 \times 4$ matrices, defined shortly.  This is a generalised
eigenvalue problem for $\sigma(k)$. For each wavenumber, it has four solutions,
whose eigenvectors~$\vect{q}(k)$ correspond, via~\eqref{eq:ansatz
1}--\eqref{eq:ansatz 4}, to disturbance amplitudes that grow exponentially with
rate $\Real(\sigma)$ and travel along the slope at wave speed $c =
-\Imag(\sigma) / k$.  Instability occurs when any of these solutions
exponentially diverges from the steady state, \ie when $\Real[\sigma(k)] > 0$.
The matrices are:
\begin{subequations}
    \begin{gather}
    \matr{A} = \begin{pmatrix}
        1 & 0 & 0 & 0 \\
        \psi_0 & 0 & 1 & 0 \\
        1 & 1 & \Delta\rho & 0 \\
        0 & 0 & 0 & 1 \\
    \end{pmatrix},\quad
    \matr{B} = \begin{pmatrix}
        1 & 1 & 0 & 0 \\
        \psi_0 & \psi_0 & 1 & 0 \\
        1 + \Fr^{-2} & 2 & \Delta\rho(1 + \frac{1}{2}\Fr^{-2}) &
        \Fr^{-2} \\
        Q_{h_0} & Q_{u_0} & Q_{\psi_0} & 0 \\
    \end{pmatrix}
    \label{eq:A and B}%
    \tag{\theequation\emph{a,b}}%
\end{gather}
\label{eq:A or B}%
and
\begin{equation}
    \matr{C} = \begin{pmatrix}
        -\Gamma_{h_0} & -\Gamma_{u_0} & -\Gamma_{\psi_0} & 0 \\
        -\Gamma_{h_0} & -\Gamma_{u_0} & -\Gamma_{\psi_0} & 0 \\
        \tau_{h_0} - 1 -\upsilon_0 \Gamma_{h_0} & \tau_{u_0} -\upsilon_0
        \Gamma_{u_0} & \tau_{\psi_0} - \Delta\rho -\upsilon_0 \Gamma_{\psi_0} & 0 \\
        \Gamma_{h_0} & \Gamma_{u_0} & \Gamma_{\psi_0} & 0 \\
    \end{pmatrix}.
    \tag{\theequation\emph{c}}%
\end{equation}
\end{subequations}
For the sake of neatness, we have used some notational shorthand to simplify
the entries.  In particular, we set $\Delta\rho \equiv \tilde\psi_b(\rho_s -
\rho_f)$, so that 
\begin{equation}
    \rho(\psi) = \rho_f + \Delta\rho \psi,\label{eq:rho}
\end{equation}
by~\eqref{eq:density} and~(\ref{eq:variable subs all}\emph{e,k,l}).  The matrices
$\matr{B}$ and $\matr{C}$ depend on linear expansions of the unknown functions
$Q$, $\tau$ and $\Gamma$ around the steady state.  In these cases, we have
written $f_{\zeta_0} \equiv \frac{\partial f}{\partial \zeta}\big|_{1,1,\psi_0}$
for each $f \in \{Q, \tau, \Gamma\}$ and $\zeta \in \{h, u, \psi, b\}$. Note
that, in deriving $\matr{B}$ and $\matr{C}$, our assumption of a homogeneous bed
allowed us to set $Q_{b_0} = \tau_{b_0} = \Gamma_{b_0} = 0$.  Finally, the basal
slip velocity, evaluated at the steady state, is denoted as $\upsilon_0 \equiv
u_b(1,1,\psi_0,b_0)$.  


\subsection{Hydraulic limit}
\label{sec:hydraulic limit}
We begin our analysis by briefly recapping the `purely hydraulic' stability
problem within our framework. That is, we address the limiting case of weak
morphodynamic processes, by sending both $Q \to 0$ and $\Gamma \to 0$.
In this case, perturbations in $\psi$ and $b$ can only be advected along the
slope, since there are no morphodynamic feedbacks through which they may grow or
decay.  Equation~\eqref{eq:morpho eigenproblem} possesses the solutions $\sigma
= -\im k$ and $\sigma = 0$, that respectively correspond to these modes of
disturbance.  The remaining two solutions are
\begin{equation}
    \sigma = -\im k - \frac{\tau_{u_0}}{2} \pm 
    \sqrt{
    \tau_{u_0}^2 / 4 - k^2 / \Fr^2
 + \im k (\tau_{h_0} - 1)
}.
    \label{eq:hydraulic sigma}
\end{equation}
These branches correspond to disturbances in the hydraulic governing equations
for $h$ and~$u$, studied in the case of general drag by \cite{Trowbridge1987}.
When $k = 0$, they pass through $\sigma = -\tau_{u_0}$ and $0$.  It can be shown
straightforwardly that $\Real(\sigma)$ is a monotonic function with respect
to~$|k|$, meaning that the maximum growth for each branch must occur at either
$k = 0$, or in the limit~$|k|\to\infty$.  Growth rate saturation at short
wavelengths is a known property of the classical roll wave instability that
highlights the omission of physics (\eg turbulent dissipation) that would
otherwise damp out disturbances over short length scales.  Evaluating the limit
of~\eqref{eq:hydraulic sigma} as $|k|\to\infty$ yields
\begin{equation}
    \Real(\sigma) \to \frac{-\tau_{u_0} \pm |1 - \tau_{h_0}|\Fr}{2}.
    \label{eq:resig lim}
\end{equation}
If $\tau_{u_0} < 0$, then there is always unstable growth (\ie at $k=0$).
However, we consider the more physically reasonable situation where $\tau_{u_0} >
0$ (\ie a drag parametrisation that increases resistance to flow at higher shear
rates). Then, if $\tau_{h_0} = 1$, both branches are everywhere stable and
asymptote to $\Real(\sigma) = -\tau_{u_0}/2$. Otherwise, since the argument of
the square root in~\eqref{eq:hydraulic sigma} always has a non-zero imaginary
part (away from $k = 0$), the growth rates are always distinct and in
particular, the branch with positive root always dominates. This turns unstable
when~\eqref{eq:resig lim} exceeds zero, which occurs if
\begin{equation}
    \Fr > \frac{\tau_{u_0}}{|1 - \tau_{h_0}|}.
    \label{eq:trowbridge criterion}%
\end{equation}
This is the stability criterion due to \cite{Trowbridge1987}, written in our
dimensionless quantities. Inclusion of the absolute value in the denominator
constitutes a minor correction to the original formula that accounts for the
case where $\tau_{h_0} > 1$. 

\subsection{Suspended load model}
\label{sec:bed exchange}%
We now reintroduce morphodynamics, by allowing for non-vanishing mass exchange
with the bed ($\Gamma \neq 0$), but continuing to neglect bed load transport ($Q
= 0$). This substantially complicates~\eqref{eq:morpho eigenproblem}, which 
becomes a fully $4\times 4$ problem. Motivated by the above discussion, we divide
our morphodynamic analysis into two tractable regimes: the long-wave (or global) limit $k = 0$ and the short-wave limit $k \gg 1$, and verify later that
these limits control most of the important aspects of the problem.

\subsubsection{Global modes: $k = 0$}
\label{sec:k = 0}%
A given steady morphodynamic flow is specified by four state variables $\tilde h_0$,
$\tilde u_0$, $\tilde \psi_0$ and $\tilde b_0$, which are constrained by only
two equations~(\ref{eq:steady morpho balance}\emph{a,b}). Therefore, the
solution space is underdetermined and there 
is a two-dimensional linear family of possible steady
states. In nature, selection of a particular flow from this family is assured
via some boundary condition, such as the total flux of material through a flow
cross-section. Moreover, transitions
from one steady flow to another within this space can occur (\eg through an increase
in the total flux).
Infinitesimal transitions between steady states are linear perturbations in
the sense of~\eqref{eq:ansatz 1}--\eqref{eq:ansatz 4}, with
$k=0$ and $\sigma = 0$ (neutral stability).
%
%
Therefore, by~\eqref{eq:morpho eigenproblem} they
satisfy $\matr{C}\vect{q} =
\vect{0}$. Solving for $\vect{q}$ reveals a two-dimensional space of
neutral modes spanned by
\begin{subequations}
\begin{gather}
    \vect{v}_1 =
    \begin{pmatrix}
        (\tau_{\psi_0} - \Delta\rho) \Gamma_{u_0} - \tau_{u_0}\Gamma_{\psi_0} \\
        (\Delta\rho - \tau_{\psi_0}) \Gamma_{h_0} + (\tau_{h_0} - 1)\Gamma_{\psi_0} \\
        \tau_{u_0}\Gamma_{h_0} - (\tau_{h_0} - 1)\Gamma_{u_0} \\
        0
    \end{pmatrix},\quad
    \vect{v}_2 = \vect{e}_4,
    \tag{\theequation\emph{a,b}}%
\end{gather}
\end{subequations}
where we adopt the convention of using $\vect{e}_j$ to denote the $j$-th
standard basis vector.  The first of these, $\vect{v}_1$, may be interpreted in
the following way.  Written in our dimensionless variables, the equations for
steady flows~(\ref{eq:steady morpho balance}\emph{a,b}) are the roots of the
function $\vect{F}(h, u, \psi) = (\tau - \rho h, \Gamma)^T$.  It is
straightforward to verify that $\nabla \vect{F}(h_0,u_0,\psi_0) \cdot \vect{v}_1
= \vect{0}$ and therefore $\vect{v}_1$ represents a shift along the curve of
solutions, implicitly defined by $\vect{F} = \vect{0}$.  The second neutral mode
$\vect{v}_2$ accounts for invariance to arbitrary translations of the bed
height.

The remaining two global modes have non-zero growth rate and therefore, by~\eqref{eq:morpho
eigenproblem}, they obey
\begin{equation}
    \sigma\matr{A}\vect{q} + \matr{C}\vect{q} =
    \vect{0}.
    \label{eq:k = 0 eigenproblem}
\end{equation}
After factoring out the neutral growth rates, the characteristic equation 
yields a quadratic from which the remaining two eigenvalues may be directly
computed. The full set of eigenvalues of~\eqref{eq:k = 0 eigenproblem} is
then
\begin{subequations}
\begin{gather}
    \sigma = 0~\mathrm{(repeated)}, \quad \frac{s_0 \pm \sqrt{s_c}}{2},
    \tag{\theequation\emph{a,b}}
\end{gather}
\label{eq:k = 0 sigma}
\end{subequations}
where $s_0$, $s_c$ are placeholders for 
\begin{subequations}
\begin{gather}
    s_0 = -\tau_{u_0} + \Gamma_{h_0} + (\upsilon_0 - \rho_b) \Gamma_{u_0} +
    (1 - \psi_0)\Gamma_{\psi_0},\label{eq:s0}\\
\begin{aligned}
    s_c = \Gamma_{u_0}^2 (\upsilon_0 - \rho_b)^2 
    + 2\Gamma_{u_0}
        \left\{
            (\upsilon_0 - \rho_b)
            \left[
                \Gamma_{h_0} + (1-\psi_0)\Gamma_{\psi_0} - \tau_{u_0}
            \right] \right.\\
            \left.
            -2\left[\tau_{h_0} + \tau_{\psi_0}(1 - \psi_0) - \rho_b\right]
        \right\}
    + \left[
        \tau_{u_0} + \Gamma_{h_0} + (1 - \psi_0)\Gamma_{\psi_0}
    \right]^2\!\!.\label{eq:sc}
\end{aligned}
\end{gather}
\label{eq:s0andsc}%
\end{subequations}
Here, we have made use of~\eqref{eq:rho} with $\psi = \psi_0$ and $\psi = \psi_b
= 1$, to eliminate $\Delta\rho$ in favour of
the bed density $\rho_b = 1 + \Delta\rho (1 - \psi_0)$ in these expressions,
which nevertheless depend on all nine independent
quantities in the matrices $\matr{A}$ and $\matr{C}$.
Before moving on to the next section, we note two important special cases.

In the non-erosive limit $\Gamma \to 0$,~\eqref{eq:s0} and~\eqref{eq:sc}
reduce to simply $s_0 = -\tau_{u_0}$ and $s_c = \tau_{u_0}^2$.
Substituting these into~(\ref{eq:k = 0 sigma}\emph{b}) leaves only one
(typically negative) non-zero growth rate, $\sigma = -\tau_{u_0}$,
consistent with the analysis in~\S\ref{sec:hydraulic limit}.

If instead, $\Gamma$ is finite, but $|\Gamma_{u_0}|$ is sufficiently small,
relative to the other components of~(\ref{eq:s0andsc}\emph{a,b}),
so that it may be neglected,
the non-zero eigenvalues become
\begin{subequations}
\begin{gather}
    \sigma = \frac{s_0 - \sqrt{s_c}}{2} = -\tau_{u_0}~~\mathrm{and}~~\sigma=
    \frac{s_0 + \sqrt{s_c}}{2} = \Gamma_{h_0} + \Gamma_{\psi_0}(1 -
    \psi_0).
    \tag{\theequation\emph{a,b}}
\end{gather}
\label{eq:approx k = 0 sigma}%
\end{subequations}
Since the latter eigenvalue (later referred to as $\sigma_a$) may be positive,
there exists a route to a purely morphodynamic instability in this case, which
depends on the signs and relative magnitudes of $\Gamma_{h_0}$ and
$\Gamma_{\psi_0}$.  Positive values for these derivatives imply positive
morphodynamic feedbacks, amplifying the flow depth and concentration
respectively. We return to this in \S\ref{sec:erosive implications}, where we
demonstrate using some generic model closures that this mode can indeed be
unstable.

\subsubsection{Short wavelengths: $k \gg 1$}
\label{sec:k gg 1}%
We now focus on short-wavelength perturbations. 
By analogy with the non-morphodynamic case of~\S\ref{sec:hydraulic limit}, we
anticipate that the limit $k\to\infty$ controls the onset and growth of
instabilities by maximising $\Real[\sigma(k)]$. (We confirm that this is often the case
for example model closures in~\S\ref{sec:erosive implications}.)
The form of~\eqref{eq:morpho eigenproblem} suggests the following asymptotic
expansions for the four growth rates and their corresponding eigenmodes in this
regime:
\begin{subequations}
    \begin{gather}
    \sigma = -\im\lambda_1 k + \lambda_0 + \lambda_{-1} k^{-1} + \ldots,
    \quad
    \vect{q} = \vect{q}_0 + \vect{q}_{-1} k^{-1} +
    \ldots
    \tag{\theequation\emph{a,b}}%
    \label{eq:sigma expansions all}%
\end{gather}
\label{eq:sigma expansions}%
\end{subequations}
Here, $\lambda_{1}$, $\lambda_0$, $\lambda_{-1}$ and $\vect{q}_0$,
$\vect{q}_{-1}$, are unknown constants and vectors to be determined shortly.
Substituting these expressions into~\eqref{eq:morpho eigenproblem} and retaining
only the leading $O(k)$ terms leaves an eigenproblem for $\lambda_1$:
\begin{equation}
    \lambda_1\matr{A}\vect{q}_0 = \matr{B}\vect{q}_0.
    \label{eq:O(k) 1}
\end{equation}
This may be solved to obtain four distinct values
\begin{equation}
    \lambda_1 = 
    1 \pm \Fr^{-1},~
    1,~
    0.
    \label{eq:lambdas}
\end{equation}
Since $c = -\Imag(\sigma) / k \to \lambda_1$ as $k\to\infty$,
these are the wave speeds for disturbances in the short-wavelength regime (and
also the characteristics of the governing equations in this context). The
corresponding eigenvectors of~\eqref{eq:O(k) 1} are
\begin{equation}
    \vect{q}_0 = 
    \begin{pmatrix}
        \pm \Fr \\
        1 \\
        0\\
        0
    \end{pmatrix},~
    \begin{pmatrix}
        \Delta \rho/2 \\
        0 \\
        -1 \\
        0
    \end{pmatrix},~
    \begin{pmatrix}
        1 \\
        -1 \\
        0\\
        \Fr^2 - 1
    \end{pmatrix}
    .
    \label{eq:O(1) eigvecs}
\end{equation}
Recalling the definition $\vect{q}=(h_1,u_1,\psi_1,b_1)^T$ and~\eqref{eq:ansatz
1}--\eqref{eq:ansatz 4}, the elements of these vectors are the leading-order
amplitudes for each mode.  Throughout the rest of the paper, we label these
modes I--IV.  Since these asymptotic vectors separate the four solution branches
of the general linear problem~\eqref{eq:morpho eigenproblem}, it will be
convenient later to use the same labels to refer to quantities at finite~$k$,
though they may not necessarily share the properties of their asymptotic
counterparts.

The first pair of modes (I,II) in~\eqref{eq:O(1) eigvecs} contain no
morphodynamic content. Indeed, they are identical to the short-wavelength modes
of the purely hydraulic problem (\S\ref{sec:hydraulic limit}), which is
guaranteed since $\matr{A}$ and $\matr{B}$ do not depend on $\Gamma$.  They
describe disturbances in $h$ and $u$, propagating at speeds $c = 1 \pm
\Fr^{-1}$. Mode~III couples unit speed perturbations in $\psi$ with the flow
free surface, while mode~IV is stationary ($c=0$) and disturbs the bedform, as
well as $h$ and $u$.

The second term in the expansion of $\sigma$ determines the leading-order real
part of the growth rate.  We substitute~\eqref{eq:sigma expansions all} back
into~\eqref{eq:morpho eigenproblem} and subtract away the $O(k)$
component, \ie\eqref{eq:O(k) 1}.
Retaining only $O(1)$ terms in the remaining equation, leaves
\begin{equation}
    \lambda_0 \matr{A}\vect{q}_0 + \im (\matr{B} - \lambda_1
    \matr{A})\vect{q}_{-1}
    = \matr{C}\vect{q}_0.
    \label{eq:O(1) 1}
\end{equation}
The unknown vector $\vect{q}_{-1}$ can be eliminated by solving the
eigenproblem adjoint to~\eqref{eq:O(k) 1}, which yields vectors $\vect{r}_0$
such that $\lambda_1 \vect{r}_0^T \matr{A} = \vect{r}_0^T \matr{B}$.
Multiplying~\eqref{eq:O(1) 1} on the left by $\vect{r}_0^T$ and rearranging
gives the formula
\begin{equation}
    \lambda_0 =
    -\frac{\vect{r_0}\cdot\matr{C}\vect{q}_0}{\vect{r}_0\cdot\matr{A}\vect{q}_0}.
    \label{eq:mu0 formula 1}
\end{equation}
Using this, the following four expressions for $\lambda_0$ are obtained, which
we label $\lambda_{0,1}, \ldots, \lambda_{0,4}$ for later reference:
\begin{subequations}
    \begin{gather}
    \lambda_{0,1} = 
    \frac{f_+(\Fr)}{\Fr(\Fr + 1)},\quad
    \lambda_{0,2} = 
    \frac{f_-(\Fr)}{\Fr(\Fr - 1)},%
    \tag{\theequation\emph{a,b}}\\
    \lambda_{0,3} = 
    (1 - \psi_0)(\Gamma_{\psi_0} - \Delta\rho \Gamma_{h_0} / 2),\quad
    \lambda_{0,4} = 
    \frac{\Gamma_{u_0} - \Gamma_{h_0}}{\Fr^2 - 1}.
    \tag{\theequation\emph{c,d}}%
\end{gather}
    \label{eq:invisc growth rates}%
\end{subequations}
By~(\ref{eq:sigma expansions}\emph{a}),
these asymptotic values dictate the limits of $\Real(\sigma)$ as $k\to
\infty$ for modes~I--IV. We list them in
the same order as their respective wave speeds
in~\eqref{eq:lambdas} and the $O(1)$ eigenvectors in~\eqref{eq:O(1) eigvecs}.
The functions $f_\pm$ are third-order polynomials in $\Fr$ defined by
\begin{align}
\begin{aligned}
    f_\pm(\Fr) = &\pm\frac{1}{2}\left[
	\Gamma_{h_0}(\upsilon_0 - \rho_b) + (1 - \tau_{h_0})
	\right] \Fr^3 \\
&+\frac{1}{2}\left[
	\Gamma_{h_0} (2\upsilon_0 - \rho_b + 1) / 2
	+ \Gamma_{u_0} (\upsilon_0 - \rho_b)
	+ 1 - \tau_{h_0} - \tau_{u_0}
	\right] \Fr^2 \\
&\pm\frac{1}{4} \left[
	\Gamma_{h_0}(\rho_b - 1) + \Gamma_{u_0} (2\upsilon_0 - \rho_b + 1) - 2\tau_{u_0}
	\right] \Fr	
	+ \frac{1}{4} 
	\Gamma_{u_0} (\rho_b - 1).
\end{aligned}
\label{eq:f pm}%
\end{align}
It is easily confirmed that as $\Gamma \to 0$, $\lambda_{0,1}$ and
$\lambda_{0,2}$ reduce to the high-$k$ growth rates of the non-erosive problem,
given in~\eqref{eq:resig lim}, while $\lambda_{0,3}, \lambda_{0,4}\to 0$.
For this reason, we will sometimes label $\lambda_{0,1}$, $\lambda_{0,2}$
and their corresponding modes~(I,II) as `hydraulic' and $\lambda_{0,3}$,
$\lambda_{0,4}$ as `morphodynamic' even though all of~(\ref{eq:invisc growth
rates}\emph{a--d}) are coupled to the
bed and sediment dynamics when $\Gamma$ is non-vanishing.

Just as in the non-erosive problem, the asymptotic growth rates
in~(\ref{eq:invisc growth rates}\emph{a}--\emph{d}) are non-zero, but typically
finite. However, there is an extra complication.  Since $f_\pm(0) = \Gamma_{u_0}
(\rho_b - 1) / 4$ and $f_\pm(\mp 1) = (\Gamma_{h_0} - \Gamma_{u_0})/2$, the
pairs $(\lambda_{0,1}, \lambda_{0,2})$ and $(\lambda_{0,2}, \lambda_{0,4})$
possess singularities at $\Fr = 0$ and $\Fr = 1$ respectively (provided
$\Gamma_{h_0} \neq \Gamma_{u_0} \neq 0$).

When $\Fr = 0$, the steady flow velocity is zero.  The singularities in the
expressions for $\lambda_{0,1}$ and $\lambda_{0,2}$ are artefacts arising from
the fact that the time scale chosen to non-dimensionalise~(\ref{eq:sw morpho
nondim}\emph{a--d}) vanishes in the limit $\tilde u_0 \to 0$. Referring back
to~(\ref{eq:variable subs all}\emph{b,h,l}), we may rewrite~(\ref{eq:invisc
growth rates}\emph{a,b}) in dimensional units and verify that these growth rates
remain finite. Specifically, when $\tilde u_0 = 0$, the expressions are
$\tilde{\lambda}_{0,j} = (3/4-j/2)\tilde \Gamma_{\tilde u_0}(\tilde\rho_b/\tilde
\rho_0 - 1)(g \cos\phi / \tilde h_0)^{1/2}$, for $j = 1,2$.

However, the singularities in~(\ref{eq:invisc growth rates}\emph{b,d}) at unit
Froude number cannot be removed by a choice of units.  They occur when the
wave speeds $\lambda_1 = 1 - \Fr^{-1}, 0$ for disturbances to the flow and
bedform, coalesce, as do the corresponding $O(1)$ modes in~\eqref{eq:O(1)
eigvecs}. Since $\lambda_0$ cannot be $O(1)$ at this singular point, our
expansions in~\eqref{eq:sigma expansions all} are inappropriate here.
Therefore, we propose instead that at $\Fr = 1$ (and for modes II~\&~IV only),
$\sigma$ and $\vect{q}$ take the asymptotic form
\begin{subequations}
\begin{gather}
    \sigma = \lambda_{1/2} k^{1/2} + \lambda_0 +
    \ldots,
    \quad
    \vect{q} = \vect{q}_0 + \vect{q}_{-1/2}k^{-1/2} + 
    \vect{q}_{-1}k^{-1} +
    \ldots,
    \tag{\theequation\emph{a,b}}%
\end{gather}
\end{subequations}
where $\lambda_{1/2}$, $\lambda_0$ and $\vect{q}_0$, $\vect{q}_{-1/2}$,
$\vect{q}_{-1}$ are to be determined. We proceed as before, by
substituting these expressions into~\eqref{eq:morpho eigenproblem} and isolating
its constituent parts at different orders in~$k$.
Retaining only~$O(k)$ terms yields $\matr{B}\vect{q}_0 = 0$, with only one
solution, $\vect{q}_0 = (1, -1, 0, 0)^T$. As it must, this matches the
coalescent modes in~\eqref{eq:O(1) eigvecs}, when they are evaluated at $\Fr =
1$.
At $O(k^{1/2})$, we have
\begin{equation}
    \lambda_{1/2}\matr{A}\vect{q}_0 + \im\matr{B}\vect{q}_{-1/2} = 0.
\end{equation}
On substituting $\vect{q}_0$ into this equation, a little algebra shows
that $\vect{e}_4 \cdot \vect{q}_{-1/2} = 2 \im \lambda_{1/2}$.
To find $\lambda_{1/2}$, we use the
$O(1)$ equation, which is
\begin{equation}
    \lambda_0 \matr{A}\vect{q}_0
    + \lambda_{1/2} \matr{A} \vect{q}_{-1/2} 
    + \im \matr{B}\vect{q}_{-1}
    + \matr{C} \vect{q}_0
    =
    \vect{0}.
    \label{eq:ksqrt O(1)}
\end{equation}
Now we notice that $\vect{e}_4 \cdot \matr{A} \vect{q}_0 = \vect{e}_4 \cdot
\matr{B} \vect{v} = 0$ for any vector $\vect{v}$. Therefore, projecting~\eqref{eq:ksqrt O(1)} onto~$\vect{e}_4$ eliminates the unknowns $\lambda_0$
and $\vect{q}_{-1}$. On doing this, substituting our expressions for
$\vect{q}_0$ and $\vect{e}_4 \cdot \vect{q}_{-1/2}$ from above and rearranging,
we find
\begin{equation}
    \lambda_{1/2} = \pm  \frac{1 + \im}{2}
    \left(\Gamma_{h_0} - \Gamma_{u_0}\right)^{1/2}.
    \label{eq:lambda half}
\end{equation}

Except for the particular case $\Gamma_{h_0} = \Gamma_{u_0}$ (where there is no
singularity in $\lambda_{0,2},\lambda_{0,4}$), these expressions have non-zero
real part.  Therefore, at $\Fr = 1$ and for $k\gg 1$, the second and fourth
modes~\eqref{eq:O(1) eigvecs} of the linear stability problem~\eqref{eq:morpho
eigenproblem} diverge with amplitudes $\sim\exp(\pm A \sqrt{k}t)$, where $A =
|\Real(\lambda_{1/2})|$. 
Crucially, one of these amplitudes is strictly positive and unbounded in the
limit $k \to \infty$. This implies that the
morphodynamic governing equations~(\ref{eq:sw morpho nondim}\emph{a--d}) at are
ill posed as an initial value problem when $\Fr = 1$, because their solutions do
not depend continuously on initial data~\citep{Joseph1990}. 
Mathematically speaking, this is a direct consequence of the model losing the
property of strict
hyperbolicity when its characteristic wave speeds~\eqref{eq:lambdas} intersect.
More intuitively, problems arise
because over any finite time interval, there are short-wavelength disturbances
that grow arbitrarily rapidly, making it impossible for 
the governing equations to behave in a physically consistent way.
This fact has practical
consequences beyond the theory of steady flows on constant slopes.  Computer
simulations of these models (in both one and two spatial dimensions) conducted on complex topographies almost inevitably
feature locations where the conditions locally match our problem at unit Froude
number and shortwave oscillations can grow catastrophically.  Numerical `solutions' in this case
may nevertheless look physically reasonable, 
since spatial discretisation imposes an upper limit on $k$. 
However, they will not converge as the numerical resolution increases
and cannot be relied upon to model real flows.

The essential issue of unbounded growth rates in these models was recognised by
\cite{Balmforth2012}, who studied the stability of a similar, but nonequivalent
system: uniform flows eroding at a constant positive rate in the Saint-Venant
equations. In the limit of slow erosion, they also observed that the
high-wavenumber growth rate of perturbations suffers a singularity at unit
Froude number.  Moreover, they were able to show that the inclusion of a
diffusive term in the momentum dynamics was sufficient to regularise their
system.
Our more general setting adds dynamic coupling with the solid phase and an
arbitrary basal drag parametrisation, thereby demonstrating that the same
problem affects a far broader range of shallow morphodynamic flow models.
Indeed, it suggests that in any situations close to, but not strictly covered by
our framework, it is important to check carefully whether the governing
equations are well posed and amend them if necessary.  Therefore, we continue
with an analysis of how this might be achieved.

\subsection{Regularisation}
\label{sec:regularisation}
We shall introduce a new term to~\eqref{eq:sw morpho nondim 3} in order to quash
unbounded growth at small length scales.  As noted by~\cite{Joseph1990}, ill
posedness often signals that there are physical processes missing from a model.
In our case, two possible culprits are the shallow-layer approximation and the
omission of bed load from the current analysis. We assess the effect of bed load
shortly, in~\S\ref{sec:bed load}.  A particular effect neglected by the
assumption of shallow flow is the aggregate loss of horizontal momentum caused
by turbulent eddies. This is usually acceptable, since it is only significant at
length scales shorter than the flow depth.  However, for short waves it
is no longer strictly negligible.  A simple and common way to include this
missing physics is to try to capture it via diffusion-like process.  We denote a
characteristic eddy viscosity for the flow by $\tilde \nu$ and non-dimensionalise
by setting $\nu = \tilde \nu / (\tilde u_0 \tilde l_0)$. 
This free parameter sets the scale of the diffusive term
$\frac{\partial~}{\partial x}(\nu \rho h \frac{\partial u}{\partial x})$, which
we add to the right-hand side of~\eqref{eq:sw morpho nondim 3}. 
We note that the extra term does not affect the steady uniform layer itself. 
Similar expressions have been employed elsewhere, as a regularisation term by
\cite{Balmforth2012} in their analysis and in the
shallow-flow models of \cite{Simpson2006,Xia2010} and \cite{Langendoen2016}.
Later, in~\S\ref{sec:effect of regularisation} we briefly address the
implications of adding a similar term to~\eqref{eq:sw morpho nondim 2} to
encapsulate turbulent sediment diffusivity.

It is unclear \emph{a priori} whether the eddy viscosity term is sufficient 
to regularise the ill-posed model equations on its own. Therefore, we must
extend the
high-wavenumber growth rate analysis of \S\ref{sec:k gg 1}.
With the extra term, the linearised system of~\eqref{eq:morpho eigenproblem}
generalises to
\begin{equation}
    \sigma\matr{A}\vect{q} + \im k \matr{B}\vect{q} + \matr{C}\vect{q} =
    -k^2 \matr{D}\vect{q}, 
    \label{eq:morpho eigenproblem visc}
\end{equation}
where $\matr{D} = (D_{ij})$ is a $4\times 4$ matrix with entries $D_{32} = \nu$
and $D_{ij} = 0$ otherwise. At high wavenumber, the leading-order component in
the linearised momentum equation is given by the new diffusive term itself.
Suppose that there is at least one eigenvalue that balances this term. This
motivates the following asymptotic expansions for $\sigma$ and $\vect{q}$ when
$k \gg 1$:
\begin{subequations}
    \begin{gather}
    \sigma = \lambda_2 k^2 + \lambda_1 k + \lambda_0 + \ldots,
    \quad
    \vect{q} = \vect{q}_0 + \vect{q}_{-1} k^{-1} + \vect{q}_{-2} k^{-2} + 
    \ldots,
    \label{eq:sigma expansions visc}%
    \tag{\theequation\emph{a,b}}%
    \end{gather}
\end{subequations}
where $\lambda_2$, $\lambda_1$, $\lambda_0$ and $\vect{q}_0$, $\vect{q}_{-1}$,
$\vect{q}_{-2}$ are to be determined.
At $O(k^2)$, equation~\eqref{eq:morpho eigenproblem visc} reduces to the
eigenproblem
\begin{equation}
\lambda_2 \matr{A} \vect{q}_0 = -\matr{D}\vect{q}_0,
\label{eq:Oksqr visc}
\end{equation}
with characteristic equation
$-\lambda_2^3(\lambda_2 + \nu) = 0$.
When $\lambda_2 = -\nu$, it may be easily verified that $\vect{q}_0 =
\vect{e}_2$.
Therefore, the diffusion operator creates one stable eigenvalue $\sigma = -\nu
k^2 + O(k)$ associated with viscous damping of $u$. The remaining three
solutions are all
$\lambda_2 = 0$ and so in these cases $\sigma$ is determined by a lower order
balance. Using~\eqref{eq:Oksqr visc}, the corresponding vector $\vect{q}_0$
is determined up to the eigenspace spanned by $\vect{e}_1$, $\vect{e}_3$ and
$\vect{e}_4$. 

We shall concentrate on the three eigenvalues with $\lambda_2 = 0$, since the
mode with $\lambda_2 = -\nu$ is always stable at high $k$.  Then, at $O(k)$,
equation~\eqref{eq:morpho eigenproblem visc} becomes
\begin{equation}
    (\lambda_1 \matr{A} + \im\matr{B})\vect{q}_0 
    =
    -\matr{D} \vect{q}_{-1}.
    \label{eq:O(k) visc}
\end{equation}
Note that 
$\vect{e}_j^T \matr{D} = \vect{0}$ for $j = 1,2,4$, since eddy viscosity appears
only in the third row of~\eqref{eq:morpho eigenproblem visc}. We can
therefore eliminate the unknown $\vect{q}_{-1}$ in~\eqref{eq:O(k)
visc} as so
\begin{equation}
    \vect{e}_j \cdot
    (\lambda_1 \matr{A} + \im\matr{B})\vect{q}_0
    =
    \vect{0},\quad\text{for}~j=1,2,4.
    \label{eq:Ok reduced visc}
\end{equation}
Likewise, we can write down the eigenproblem adjoint to~\eqref{eq:O(k) visc} as
$(\lambda_1\matr{A} + \im\matr{B})^T\vect{r}_0 = -\matr{D}^T\vect{r}_{-1}$,
where $\vect{r}_0$ and $\vect{r}_{-1}$ are unknown left eigenvectors. 
We will shortly need the constrained problem, with
$\vect{r}_{-1}$ eliminated as so
%
\begin{equation}
    \vect{e}_k \cdot
    (\lambda_1 \matr{A} + \im\matr{B})^T\vect{r}_0
    =
    \vect{0},\quad\text{for}~k=1,3,4.
    \label{eq:Ok reduced visc adjoint}
\end{equation}

Expanding $\vect{q}_0 = a_1\vect{e}_1 + a_3 \vect{e}_3 + a_4 \vect{e}_4$
in~\eqref{eq:Ok reduced visc} results in a $3\times3$ eigenvalue problem for the
unknowns $a_1$, $a_3$, $a_4$ and $\lambda_1$. Its characteristic equation is
\begin{equation}
    \lambda_1 (\lambda_1 + \im)^2 = 0.
\end{equation}
So $\lambda_1 = 0$ or $-\im$. In either case, we are forced to proceed further
to determine the leading real part of $\sigma$.
Therefore, we use the $O(1)$ part of~\eqref{eq:morpho eigenproblem visc}, which
is (for $\lambda_2 = 0$):
\begin{equation}
    (\lambda_0 \matr{A} + \matr{C})\vect{q}_0 + (\lambda_1 \matr{A} + \im \matr{B})\vect{q}_{-1}
    =
    -\matr{D} \vect{q}_{-2}.
    \label{eq:O(1) visc}
\end{equation}
We shall divide our pursuit of $\lambda_0$ according to the value of
$\lambda_1$.

\begin{enumerate}
    \item{Case: $\lambda_1 = 0$.}
Solving~\eqref{eq:Ok reduced visc} for the eigenvector yields $\vect{q}_0 =
\vect{e}_4$. 
To eliminate, the unknown vectors $\vect{q}_{-1}$ and $\vect{q}_{-2}$,
from~\eqref{eq:O(1) visc}, we simply note that $\vect{e}_4^T\matr{B} =
\vect{0}$, since the last row of $\matr{B}$ is all zeros (when $Q=0$).  
Therefore, we
project~\eqref{eq:O(1) visc} onto $\vect{e}_4$ and substitute in
$\vect{q}_0$ to give
\begin{equation}
    \lambda_0 = -\frac{\vect{e}_4\cdot \matr{C} \vect{e}_4}{\vect{e}_4 \cdot
    \matr{A}\vect{e}_4}
    = 0.
\end{equation}
Referring back to our expansions~\eqref{eq:sigma expansions visc},
this means that
there always exists an eigenpair $(\sigma, \vect{q})$ with $\sigma \to
0$ and $\vect{q} \to \vect{e}_4$ as $k\to \infty$. Note that this
situation corresponds to perturbations of the bedform.

\item{Case: $\lambda_1 = -\im$.}
Since this is a repeated root,~\eqref{eq:Ok reduced visc} only determines
$\vect{q}_0$ within a two-dimensional subspace. Straightforward algebra
gives this simply as $\vect{q}_0 = a_1\vect{e}_1 + a_3 \vect{e}_3$. Similarly,
the
corresponding adjoint eigenproblem~\eqref{eq:Ok reduced visc adjoint},
constrains the left eigenvector $\vect{r}_0$ to lie in the subspace
spanned by $\vect{e}_1$ and $\vect{e}_2$.
We project~\eqref{eq:O(1)
visc} onto these vectors, yielding
\begin{subequations}
\begin{gather}
    \vect{e}_1 \cdot (\lambda_0 \matr{A} + \matr{C})\vect{q}_0
    + \im \vect{e}_2 \cdot \vect{q}_{-1} = 0,\\
    \vect{e}_2 \cdot (\lambda_0 \matr{A} + \matr{C})\vect{q}_0
    + \im \psi_0\vect{e}_2 \cdot \vect{q}_{-1} = 0.
\end{gather}%
\label{eq:uj projection}%
\end{subequations}
Note that only the second element of $\vect{q}_{-1}$ appears in these equations.
To eliminate it, we return to the full $O(k)$ problem. Since
$\vect{e}_3\cdot\matr{D}\vect{v} = \nu \vect{e}_2 \cdot \vect{v}$ for any vector
$\vect{v}$, we project~\eqref{eq:O(k) visc} onto $\vect{e}_3$ and rearrange
to give $\vect{e}_2 \cdot \vect{q}_{-1} = -\im (2a_1 + \Delta\rho a_3) /
(2\nu\Fr^2)$. Substituting this expression into~(\ref{eq:uj
projection}\emph{a,b})
yields a $2\times 2$ system for $a_1$, $a_3$ and $\lambda_0$,
which we solve to find the two eigenvalues
\begin{equation}
    \lambda_0 = \lambda_\pm(\Fr) \equiv \frac{R}{2}
    +
    \frac{\pm\sqrt{S} - 1}{2\nu\Fr^2},
    \label{eq:lambda pm}
\end{equation}
where $R \equiv \Gamma_{h_0} + \Gamma_{\psi_0}(1-\psi_0)$ and $S \equiv (R\nu \Fr^2 + 1)^2 - 2\nu \Fr^2 \Gamma_{h_0}(\rho_b + 1)$.
One of the pair, $\lambda_-$, possesses a singularity at $\Fr = 0$.  However, as
in \S\ref{sec:k gg 1}, this is merely an artefact of our choice of
dimensionless units that may be removed by an appropriate rescaling.
The corresponding eigenvectors are
\begin{equation}
    \vect{q}_0 = \vect{e}_1 + \frac{\nu \Fr^2 (R - 2\Gamma_{h_0}) + 1 \pm
    \sqrt{S}}{2\Gamma_{\psi_0}\nu\Fr^2 - \Delta\rho}\vect{e}_3.
    \label{eq:a1 a3}%
\end{equation}
In the limit of vanishing eddy viscosity ($\nu \to 0$), $\vect{q}_0 \to
\vect{e}_1 - (1\pm 1) \vect{e}_3 / \Delta\rho$. Therefore, since we anticipate
small $\nu$, $\lambda_-$ corresponds largely to growth in $h$ only, whereas
$\lambda_+$ corresponds to coupled growth in $h$ and $\psi$.  By comparing with
the modes of the unregularised problem in~\eqref{eq:O(1) eigvecs}, $\lambda_-$
may be traced to the hydraulic modes and $\lambda_+$ to the third mode related
to solid fraction perturbations. The former is responsible for very strong
damping, since $\lambda_- \approx - 1/\nu\Fr^2$ for small $\nu$.  Conversely,
using l'H\^{o}pital's rule, it may further be verified that $\lim_{\nu\to 0}
\lambda_+ = \lambda_{0,3}$ from~(\ref{eq:invisc growth rates}\emph{c}).
Therefore, in the limit of small $\nu$, this mode is not affected by the
regularisation term.
\end{enumerate}

To recap, using~\eqref{eq:sigma expansions visc}--\eqref{eq:a1 a3}, we have
computed the growth rates $\sigma$ of perturbations for non-zero values of $\nu$
at leading-order for large wavenumber. These expressions are valid for all~$\Fr
> 0$ and models of the form~\eqref{eq:morpho eigenproblem visc}. They are
\begin{subequations}
    \begin{gather}
    \sigma = -\nu k^2 + O(k), \quad
    0 + O(k^{-1}), \quad
    -\im k + \lambda_\pm(\Fr) + O(k^{-1}),
    \label{eq:visc growth rates}%
    \tag{\theequation\emph{a--c}}%
\end{gather}
\label{eq:visc growth rates all}%
\end{subequations}
where $\lambda_\pm(\Fr)$ was defined in~\eqref{eq:lambda pm}. The corresponding
mode amplitudes are given by $\vect{q}_0 = \vect{e}_2$, $\vect{e}_4$ and the
two expressions in~\eqref{eq:a1 a3}. 
Since the
growth rates are all bounded above, we conclude that the inclusion of a
diffusive term in~\eqref{eq:sw morpho nondim 3} successfully regularises the
singularities in~\eqref{eq:morpho eigenproblem} that are otherwise present at
$\Fr = 1$, removing the problem of ill posedness.
However, since the real parts $\lambda_\pm$ of the
last two modes remain non-zero, they are still potentially unstable in the $k\gg 1$
regime
(if $\lambda_\pm > 0$). We return to this point in \S\ref{sec:erosive
implications}.

As $k\to 0$, the diffusive term vanishes in the linearised
equations~\eqref{eq:morpho eigenproblem visc} to leading-order. Therefore, the
coefficient $\nu$ sets the effective length scale over which eddy viscosity
damps out perturbations. For large-scale geophysical flows where $\nu$ is
relatively small, we may thus anticipate linear growth rates for the most part
matching those of the $\nu = 0$ problem, with eddy viscosity only affecting very
short wavelengths.  In this case, the linear stability may still be controlled
by the values of the $\nu = 0$ asymptotic growth rates given in~\eqref{eq:invisc
growth rates}. The intuition here is that for sufficiently small $\nu$ there
must be a scale separation between the `asymptotic' ($k\gg 1$) regime of the
$\nu = 0$ case and any damping (at still higher $k$) of $\sigma$ induced by
turbulent momentum diffusion. We verify this for illustrative model closures
in~\S\ref{sec:effect of regularisation}.

\subsection{Bed load}
\label{sec:bed load}%
Returning to the original system~(\ref{eq:sw morpho nondim}\emph{a--d}), with
the eddy viscosity regularisation~$\nu = 0$, we widen our perspective, to allow
for non-zero bed flux~$Q$. Unfortunately, in this case,
analytical solutions of the linear system~\eqref{eq:morpho eigenproblem} become
too complex to work with (even in the long- and short-wavelength regimes) and
cease to be useful.  Therefore, in this subsection we limit our scope to one
important concern: how is the well-posedness of the model affected by~$Q$?  To
address this, we compute the system characteristics, $\lambda_1$, as given by
solutions to~\eqref{eq:O(k) 1}, since these are sufficient to determine whether
equations~(\ref{eq:sw morpho nondim}\emph{a--d}) may be correctly posed as an
initial value problem.  Specifically, if the characteristics are all real-valued
and distinct, the system is strictly hyperbolic and well posed.  If instead, any
of the characteristics have non-zero imaginary part, the system is not hyperbolic
and ill posed~\citep[see e.g.][]{Ivrii1974}. Alternatively, if the
characteristics are all real, but one or more are repeated, the equations are
hyperbolic, but may still be ill posed, as we saw in~\S\ref{sec:k gg 1}.

In the case of bed load models ($\Gamma = 0$), strict hyperbolicity has
previously been demonstrated for various cases. A number of earlier papers
derived formulae for the system characteristics by expanding~$\lambda_1$ in
terms of parameters that are small when the bed dynamics is slow compared with
the hydraulic variables~\citep{Lyn1987,Zanre1994,Lyn2002,Lanzoni2006}.  From
this perspective, the degeneracy of the system characteristics that underpins
ill posedness in the $Q\to0$ limit is already well appreciated, since it causes
naive asymptotic formulae for $\lambda_1$ to break down near $\Fr =
1$~\citep{Lyn1987,Zanre1994}.  Later, \cite{Cordier2011} derived general
requirements for bed load models to be strictly hyperbolic. We extend their
analysis to our setting, which allows for the bulk density variations that may
arise with a suspended load.
Indeed, since $\Gamma$ does not appear in~\eqref{eq:O(k) 1},
it does not affect the characteristics and so the following analysis
applies equally well whether or not bulk entrainment is included. 

Bed load terms are most often employed in dilute systems, where we would not
expect the basal stresses that drive bed load transport to be sensitive to small
changes in the bulk solid fraction.
Therefore, we make the additional simplifying assumption
that $Q_{\psi_0}$ is small enough that it may be neglected, so $Q_{\psi_0} = 0$
in~(\ref{eq:A or B}\emph{b}). 
Then, the characteristic equation resulting from~\eqref{eq:O(k)
1} reduces to
\begin{equation}
(\lambda_1 - 1)p(\lambda_1) = 0,
    \label{eq:characteristics quartic}
\end{equation}
where
    $p(\lambda_1) = 
    \Fr^2\lambda_1^3 - 2\Fr^2\lambda_1^2 + (\Fr^2 - Q_{u_0} - 1)\lambda_1 + Q_{u_0} -
    Q_{h_0}$.
The system is strictly hyperbolic if and only if~\eqref{eq:characteristics
quartic} has four distinct real solutions.  Note that these solutions only
depend on $Q_{h_0}$, $Q_{u_0}$ and $\Fr$.  One of them, arising from the solid
mass transport equation~\eqref{eq:sw morpho nondim 2}, is always $\lambda_1 =
1$. In the particular case where $Q_{h_0} = -1$, we also have $p(1)=0$, so this
eigenvalue is degenerate.  Otherwise, if $Q_{h_0} \neq -1$, then $p(1) =
-Q_{h_0} - 1 \neq 0$, and the remaining solutions to~\eqref{eq:characteristics
quartic} are never unity.  Therefore, we only need to assess the roots of the
cubic polynomial $p$ to see if all four characteristics are distinct.

On differentiating $p$ (with respect to $\lambda_1$), its turning points may be
found at
\begin{equation}
    \lambda_1 = \frac{2}{3} \pm \frac{1}{3\Fr}\sqrt{\Fr^2 + 3(Q_{u_0} + 1)}.
    \label{eq:turning pts}%
\end{equation}
Hence, a necessary condition for strict hyperbolicity is $Q_{u_0} > -1 -\Fr^2 /
3$.  Labelling the two turning points as $\lambda_1 = \ell_\pm$, it follows
immediately from considering $p$ as $\lambda_1\to\pm\infty$ in this case,
that $\ell_-$ is always a local maximum and $\ell_+$
a local minimum. Therefore, for $p$ to possess three real roots, it must
additionally satisfy $p(\ell_-) > 0$ and $p(\ell_+) < 0$.  By evaluating $p(\ell_\pm)$
and noting that $\partial p / \partial Q_{h_0} = -1$, it is straightforward to
show that $p(\ell_-) > 0$ when $Q_{h_0} < G_+$ and $p(\ell_+) < 0$ when $Q_{h_0} >
G_-$, where
\begin{equation}
    G_\pm(\Fr, Q_{u_0}) = \frac{1}{27\Fr}\left[
    2\Fr^3 \pm 2(\Fr^2 + 3Q_{u_0} + 3)^{3/2} + 9 \Fr(Q_{u_0} - 2)
\right].
\label{eq:G pm}%
\end{equation}
Therefore, the region of parameter space where the system is strictly hyperbolic
satisfies $G_- < Q_{h_0} < G_+$. Conversely, when either $Q_{h_0} < G_-$ or
$Q_{h_0} > G_+$ two of the characteristics, \ie roots
of~\eqref{eq:characteristics quartic}, are complex conjugate and the system is
non-hyperbolic.  We now seek to identify constraints on $Q_{h_0}$ and $Q_{u_0}$
such that the system is strictly hyperbolic for all~$\Fr$.

In the particular case when $Q_{h_0} = Q_{u_0}$,
the solutions of~\eqref{eq:characteristics quartic} may
be readily computed to be 
\begin{equation}
\lambda_1 =
1 \pm \Fr^{-1} \sqrt{Q_{h_0} + 1}, 1, 0.
    \label{eq:lambdas bedload}%
\end{equation}
These expressions are commensurate with the case $Q\to 0$, whose
characteristics were given in~\eqref{eq:lambdas}. Here, only the hydraulic modes
are altered by the bed load term. All four values are distinct (so $G_- <
Q_{h_0} < G_+$), unless $\Fr =
\sqrt{Q_{h_0} + 1}$, where a hydraulic mode intersects with the bed
characteristic $\lambda_1 = 0$, and $Q_{h_0} = Q_{u_0} = G_+$. On
differentiating~\eqref{eq:G pm} with respect to $\Fr$, it can be shown that this
point is a global minimum of $G_+(\Fr, Q_{u_0})$ for any fixed $Q_{u_0}$.
Therefore, $Q_{h_0} < G_+$ for all $\Fr$ only if $Q_{h_0} < Q_{u_0}$.

\begin{figure}
    \centering
    \includegraphics{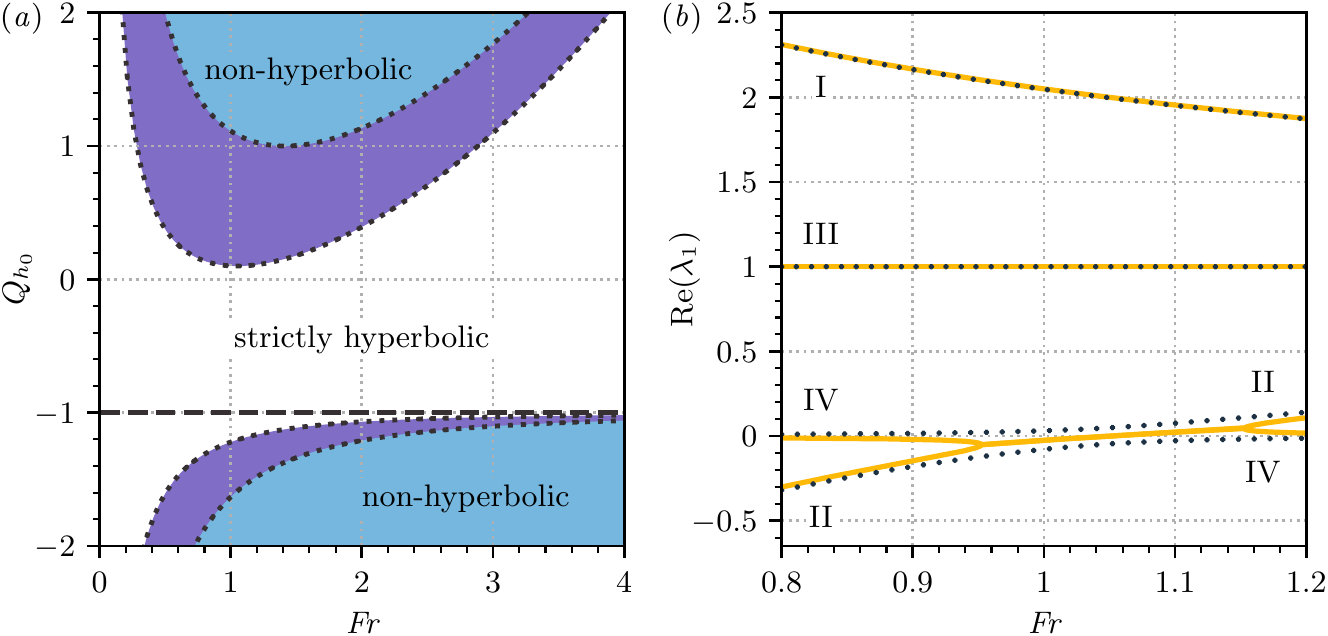}%
    \caption{%
    Hyperbolicity of the morphodynamic model equations depends on the bed
    load function $Q$.
    (\emph{a})~Regions of non-hyperbolicity as a function of $\Fr$ and
    $Q_{h_0}$, for fixed $Q_{u_0} = 0.1$~(purple shading), $1$~(blue
    shading) and~$Q_{\psi_0} = 0$.  Outside these regions, the model is strictly hyperbolic, save
    along the bounding curves~$G_\pm$~(dotted lines) and the special case
    $Q_{h_0} = -1$~(dashed line), where one of the roots of $c$ intersects
    with the solid mass transport characteristic.
    (\emph{b})~System characteristics as a function of $\Fr$ for $Q_{u_0} =
    0.1$, $Q_{\psi_0} = 0$
    and $Q_{h_0} = 0.095$ (dotted lines), $0.105$ 
    (solid lines). We label the curves I--IV according to the ordering of the
    corresponding characteristics derived in~\eqref{eq:lambdas} for the $Q=0$
    case.
}
\label{fig:Qh > Qu}
\end{figure}

The lower limit $G_-$, is a strictly increasing function of $\Fr$ (for any
$Q_{u_0}$), with $G_-(\Fr, Q_{u_0}) \to -\infty$ as $\Fr \to 0$ and $G_-(\Fr,
Q_{u_0}) \to -1$ as $\Fr \to \infty$.  Combining this information with the lower
bound for $G_+$, we conclude that the system (with $Q_{\psi_0}=0$ assumed) is
strictly hyperbolic over all Froude numbers if and only if
\begin{equation}
-1 < Q_{h_0} < Q_{u_0}.
\label{eq:hyperbolic ineq}%
\end{equation}
We note, with reference to~\eqref{eq:turning pts}, that this stronger condition automatically satisfies the requirement that
$p$ has two turning points.
In figure~\ref{fig:Qh > Qu}(\emph{a}), we plot examples of the bounds $G_\pm$,
as (dotted) curves in $(\Fr,Q_{h_0})$-space for fixed $Q_{u_0}$, indicating the regions
where the model fails to be hyperbolic.
The axes have been chosen so as to encompass very general $Q$ closures. However,
fortunately in applications sediment flux typically depends only very weakly, if
at all on the flow depth, so $|Q_{h_0}| \ll 1$ is expected.  In fact, it is
common in fluvial models to have $Q_{u_0}$ strictly positive and $Q_{h_0} =
0$ or $-1 \ll Q_{h_0} < 0$ (\eg in the latter case, if a Manning friction law is
employed).  Therefore, most studies that include bed load operate in a regime
where~\eqref{eq:hyperbolic ineq} is satisfied.

This analysis suggests an alternative to the regularisation strategy of
\S\ref{sec:regularisation}, since adding even a small bed load flux term can
ensure that the model equations are well posed, provided
that~\eqref{eq:hyperbolic ineq} is satisfied.  We visualise the effect of bed
load on the system characteristics in figure~\ref{fig:Qh > Qu}(\emph{b}), either
side of the threshold $Q_{h_0} = Q_{u_0}$ where the system loses hyperbolicity.
We plot $\Real(\lambda_1)$ as a function of $\Fr$ for $Q_{u_0} = 0.1$ and
$Q_{h_0} = 0.095$ (dotted lines), $0.105$ (solid lines).  The four branches of
$\lambda_1$ are labelled I--IV, according to the ordering of the corresponding
modes in the $Q = 0$ case, adopted in~\S\ref{sec:k gg 1}. When $Q_{h_0} = Q_{u_0}
= 0.1$ (not shown), the mode~II and~IV characteristics intersect at a single
point, $\Fr = \sqrt{1.1} \approx 1.05$ in this case, which is easily calculated
using the expressions in~\eqref{eq:lambdas bedload}.
Decreasing $Q_{u_0}$ (so that $Q_{u_0} < Q_{h_0}$) causes these characteristics
to coalesce into a complex conjugate pair, resulting in a region ($0.95 \lesssim
\Fr \lesssim 1.15$) where the system is non-hyperbolic. Conversely, increasing
$Q_{u_0}$ separates the intersecting characteristics so that four real values
are present across all Froude numbers.  (Note that these separated curves are
labelled with both II and~IV, since they originate from different modes at
either end.)
The other two characteristics (I and~III) are essentially unaffected by small
changes in the bed load.

\section{Implications}
\label{sec:erosive implications}
In this section, we examine the above analyses in greater detail by choosing
some closures for the morphodynamic model equations~(\ref{eq:sw morpho
nondim}\emph{a--d}). We focus initially on the primary case of the basic
suspended load model, before investigating the effects of incorporating eddy
viscosity and bed load.
It is our
contention that the specifics of individual modelling terms should not
qualitatively affect the observations below, provided that they are
consistent with the essential physics of the problem. Therefore, in this
exposition, we favour simple phenomenological formulae.

\subsection{Model closures}
\label{sec:closures}%
In order to capture a range of different sedimentary flows, from dilute
suspensions to fully granular flow, we use a mixed drag formulation that
depends on the bulk solid fraction, writing
\begin{equation}
    \tilde \tau = (1 - \psi) \tilde \tau_f + \psi \tilde \tau_g,
    \label{eq:mixed drag}%
\end{equation}
where $\tilde \tau_f$ and $\tilde \tau_g$ are fluid and granular drag laws
respectively.
We assume that the bed is a saturated mixture of fluid and sediment
containing the maximum possible sediment concentration $\tilde \psi_b = \tilde
\psi^*$. The maximum solid fraction $\tilde \psi^*$ depends on how efficiently
particles can be packed and is typically observed to be around
$60$--$70\%$~\citep{Santiso2002,Farr2009}.
Since $\psi = \tilde \psi /
\tilde \psi_b$, we have $0 \leq \psi \leq 1$ for all flows and
therefore~\eqref{eq:mixed drag} contains all weighted combinations of fluid and
granular drag. For the fluid law, we employ the common Ch\'ezy formula for
turbulent shear stress, $\tilde \tau_f = C_d \tilde \rho \tilde u^2$, where
$C_d$ is a drag coefficient, which we assume to be constant.
For the granular drag, we use the frictional law due to \cite{Pouliquen2002},
which sets $\tilde \tau_g = \mu(I) \tilde \rho g \cos(\phi) \tilde h$. The
phenomenological parameter $\mu$ is modelled as an 
increasing function of the dimensionless
\emph{inertial number} $I \equiv uh^{-3/2}d\Fr$, 
where $d = \tilde d / \tilde h_0$ and $\tilde d$ denotes the characteristic
diameter of the sediment particles. 
It is constructed so as to vary smoothly between a lower (static) limit $\mu_1$
and an upper (dynamic) bound $\mu_2$,
with $0 < \mu_1 < \mu_2$, and takes the form
\begin{equation}
    \mu(I) = \mu_1 + \frac{\mu_2 - \mu_1}{1 + \beta I^{-1}},
    \label{eq:pouliquen mu}
\end{equation}
where $\beta$ is a positive constant that may be determined
experimentally. 

We suppose that mass transfer is governed by the competing processes of bed
erosion at a rate $\tilde E$ and particle deposition at rate $\tilde D$, writing
$\tilde \Gamma = \tilde E - \tilde D$. Since these processes take place at the
scale of individual particles, we opt to non-dimensionalise these closure terms
using the velocity $\tilde u_p = (g'_\perp \tilde d)^{1/2}$, where $g_\perp' = g
\cos\phi \left( \tilde\rho_s / \tilde\rho_f - 1 \right)$ is the reduced gravity
for a particle in dilute suspension, resolved perpendicular to the slope. The
dimensionless transfer rates are then~$E_p = \tilde E / \tilde u_p$ and~$D_p =
\tilde D / \tilde u_p$. This rescaling allows us to fix appropriate
dimensionless constants for these closures when considering the steady balance
$E_p = D_p$ independently. However, note that care must be taken when reintroducing
these terms into~(\ref{eq:sw morpho nondim}\emph{a--d}), which uses a
different velocity scale for $\partial b / \partial t$.
Specifically, if $\Gamma_p \equiv E_p - D_p$, then~(\ref{eq:variable subs
all}\emph{h}) implies that $\Gamma = \Gamma_p \Fr
d^{1/2}(\rho_s/\rho_f-1)^{1/2}\cot\phi$.

Entrainment of particles into the flow is caused by turbulent shear stresses at
the bed, which must overcome the static friction experienced by resting grains.
Competition between $\sim\tilde\tau \tilde{d}^2$ drag forces and
$\sim \tilde \rho_f g_\perp' \tilde{d}^3$ frictional forces
(assumed proportional to the submerged weight of individual grains)
can be captured by their ratio, the dimensionless \emph{Shields number}
$\theta \equiv \tilde\tau / (\tilde\rho_f g_\perp' \tilde d)$. 
Experimental observations for dilute flows suggest that at sufficiently high
drag, flow erosion obeys a power law of the form
$\tilde E \propto (\theta - \theta_c)^{3/2}$, where~$\theta_c$ is a critical
Shields number below which there is no entrainment.
Beyond this there is considerable disagreement concerning both the exact functional
form for $\tilde E$ and its magnitude~\citep{Lajeunesse2010}. Moreover, it is
unclear whether this general erosion model applies for more concentrated
suspensions, where effects such as the pore water pressure modify the force
relationship encapsulated in the Shields number. Since our aim here is only to
elucidate some general properties of solutions, we prefer simplicity here and
suppose that $\tilde E$ depends linearly on~$\tilde u_p (\theta -
\theta_c)^{3/2}$. However, we shall make one important modification to this
dilute erosion law.  Since concentrated layers may be held static on shallow
grades by their granular friction, we must not permit erosion to occur in
situations where $\theta > \theta_c$, yet $\tilde u = 0$.  Therefore, we
set~$\theta_c = \theta_c^* + \theta_0$, where~$\theta_c^*$ denotes the usual
critical Shields number (for dilute suspensions) and~$\theta_0(\tilde h, \tilde
\psi) = \theta|_{\tilde{u} = 0}$, \ie the Shields number of a resting flow,
which may become large as $\tilde\psi$ increases.  For simplicity, we consider
$\theta_c^*$ constant in this study, even though in principle it depends on flow
properties such as the particle Reynolds number $\Rey_p \equiv \tilde{u}_p
\tilde{d}/\tilde{\nu}_f$, where $\tilde \nu_f$ is the kinematic viscosity of the
fluid~\citep{Soulsby1997}.  On dividing through by $\tilde u_p$, the
dimensionless entrainment rate is then
\begin{equation}
    E_p(h, u, \psi) = \begin{cases}
        \varepsilon
        \left[\theta(h,u,\psi) - \theta_c(h, \psi)\right]^{3/2} & \text{if }
        \theta > \theta_c, \\
        0 & \text{otherwise},
    \end{cases}
    \label{eq:erosion phenom}
\end{equation}
where~$\varepsilon$ is a proportionality coefficient that characterises
the erodibility of the bed. 

We treat sediment deposition as being governed by a process of hindered
settling.  At low concentrations, particles settle independently, so the
(monodisperse) sediment deposits at a rate~$\sim \tilde w_s \psi$, where $\tilde
w_s$ denotes the characteristic falling speed of the
grains. As concentrations increase, pure settling becomes disrupted as particles
increasingly interact, ultimately shutting off as $\psi \to 1$ and grains can no
longer fall (in a time-averaged sense) under gravity. Therefore, we take the
deposition term to be
\begin{equation}
    D_p(\psi) = w_s \psi (1 - \psi),
    \label{eq:deposition phenom}
\end{equation}
where $w_s = \tilde w_s / \tilde u_p$.  This a slight simplification of the
widely used formula due to \cite{Richardson1954}.  More detailed and accurate
expressions for $D_p$ typically involve empirical fits featuring the same
essential form~\citep[e.g.][]{Spearman2017}.

When considering a bed load, we use the following standard expression, which
mirrors the entrainment rate of~\eqref{eq:erosion phenom}:
\begin{equation}
    Q_p = \begin{cases}
        \gamma
        \left[\theta(h,u,\psi) - \theta_c(h, \psi)\right]^{3/2} & \text{if }
        \theta > \theta_c, \\
        0 & \text{otherwise},
    \end{cases}
    \label{eq:mpm}%
\end{equation}
where $Q_p$ is a particle-scale non-dimensionalisation such that 
$Q_p = \tilde Q / (\tilde d \tilde u_p)$ and $\gamma$ is a constant that
dictates the flux strength. For example, $\gamma = 8$ sets the well-known
Meyer-Peter \& M\"uller formula \citep{Meyer1948}.
The corresponding flow-scale non-dimensionalisation, as used in~\eqref{eq:sw
morpho nondim 4} and given by~(\ref{eq:variable subs all}\emph{i}), is $Q = Q_p \tilde d \tilde
u_p / (\tilde h_0 \tilde u_0) = Q_p d^{3/2} (\rho_s / \rho_f - 1)^{1/2} / \Fr$.

Finally, the basal velocity closure dictates the characteristic downslope flow
speed at the bed, during particle entrainment.
Where needed, we assume that it can be modelled by a
turbulent friction velocity of the form
\begin{equation}
\tilde u_b = \sqrt{\tilde \tau / \tilde \rho}.
    \label{eq:ub}%
\end{equation}

A large number of free parameters are involved in specifying these
various model closures.  Therefore, we choose to fix some illustrative 
values, as listed in table~\ref{tab:illustrative params}, making it clear
whenever we deviate from these defaults.
\begin{table}
  \begin{center}
\def~{\hphantom{0}}
  \begin{tabular}{ccccccccccc}
      $\tilde \rho_s/\tilde \rho_f$ & $\tilde \psi_b$ & $C_d$ & $d$ & $\mu_1$ &
      $\mu_2$ & $\beta$ & $\varepsilon$ & $\theta_c^*$ & $w_s$ \\
      $2$ & $0.65$ & $0.01$ & $0.01$ & $0.1$ & $0.4$ & $0.1$ & $0.01$ & $0.05$ &
      $1$ \\
  \end{tabular}
  \caption{Illustrative model parameters. Except where otherwise stated, all
  results in \S\ref{sec:erosive implications} assume these values. Definitions
  for these parameters may be found in equations \eqref{eq:density},
  \eqref{eq:sw ero solid mass}, \eqref{eq:mixed drag}--\eqref{eq:deposition
  phenom} and the accompanying text in each case.}
  \label{tab:illustrative params}
  \end{center}
\end{table}
An investigation of other parameter choices indicated that our observations
below are qualitatively robust to variations in these values.

\subsection{Existence of steady layers}
\label{sec:existence}%
We are now in a position to assess when uniform flowing layers can exist in
equilibrium.  This is dictated by the steady balances in~(\ref{eq:steady morpho
balance}\emph{a,b}), which we recall enforce that drag balances the downslope
component of gravitational forcing and that there is no net material transfer
between the flow and the bed. Note that these conditions are independent of
whether there is a bed load or not. To begin with, we concentrate on the stress
balance and assume that mass transfer with the bulk is negligible ($\Gamma \to
0$), thereby automatically satisfying~(\ref{eq:steady morpho balance}\emph{b}).
Substituting our mixed drag law~\eqref{eq:mixed drag} into~(\ref{eq:steady
morpho balance}\emph{a}), non-dimensionalising and rearranging gives the
following condition for existence of a steady layer of solid fraction $\psi_0$:
\begin{equation}
    \tan\phi - (1-\psi_0)C_d\Fr^2 = \psi_0 \mu(d\Fr).
    \label{eq:steady drag phenom}
\end{equation}
In the dilute limit, $\psi_0 \to 0$, where drag is purely fluid-like, this
equation selects a unique Froude number, $\Fr = \sqrt{\tan\phi/C_d}$.  Such
states become unstable when $\Fr > 2$~\citep{Jeffreys1925}.  Conversely, in the
concentrated limit, $\psi_0 \to 1$, where drag is purely granular, steady layers
adopt a unique inertial number, given by $I_0 \equiv d\Fr = \mu^{-1}(\tan
\phi)$. Since $\mu_1 < \mu(I) < \mu_2$, only a range of slope angles (between
$\arctan \mu_1$ and $\arctan \mu_2$) are permitted.  The threshold for linear
instability in this case does not depend on $\mu$ and was computed
by \cite{Forterre2003} to be $\Fr > 2/3$.

For intermediate values of $\psi_0$,
since the left-hand hand side in~\eqref{eq:steady drag phenom} is a decreasing
function of $\Fr$ and unbounded below, the steady drag balance may be satisfied
as long as $\tan\phi \geq \psi_0 \mu_1$.  The effect of the Ch\'{e}zy drag term
thereby relaxes the limits on existence imposed by the granular law. Steady
layers that are more dilute can exist in mobile equilibrium at shallower slope
angles, \ie down to $\arctan (\psi_0 \mu_1)$, while steep steady flows
($\tan\phi \to \infty$) may always be maintained at a suitably high $\Fr$, since
the turbulent drag, $C_d\Fr^2$, is not bounded above. However, note that such
solutions are not necessarily stable. Indeed, the stability threshold for these
flows may be readily computed using Trowbridge's criterion~\eqref{eq:trowbridge
criterion}.
After non-dimensionalising~\eqref{eq:mixed drag} and differentiating, one sees
that $\tau_{u_0} = [2(1-\psi_0)C_d\Fr^2 + \psi_0\mu' I_{u_0}]\cot\phi$,
where 
$\mu' = \partial \mu / \partial I |_{I=I_0}$
and 
$I_{u_0}
= \partial I / \partial u |_{h,u=1} = d\Fr$.
Likewise, $\tau_{h_0} = \psi_0[\mu(d\Fr) + \mu'(d\Fr)I_{h_0}]\cot\phi$, with
$I_{h_0} = \partial I / \partial h |_{h,u=1} = -\frac{3}{2} d\Fr$.
On substituting these expressions into~\eqref{eq:trowbridge
criterion} and rearranging, one sees that these states are unstable when
\begin{equation}
    (1-\psi_0)C_d (\Fr - 2) + \psi_0 d \mu' (3/2 - 1 / \Fr) > 0.
    \label{eq:crit Fr eqn}
\end{equation}
Note that this criterion smoothly interpolates between the $\Fr = 2$ and $\Fr =
2/3$ thresholds for the special cases of purely fluid ($\psi_0 = 0$) and
granular ($\psi_0 = 1$) flows.

We summarise the existence of non-erosive layers subject to the drag
law~\eqref{eq:mixed drag}
in figure~\ref{fig:growth
rates}. 
\begin{figure}
 \centering%
 \includegraphics{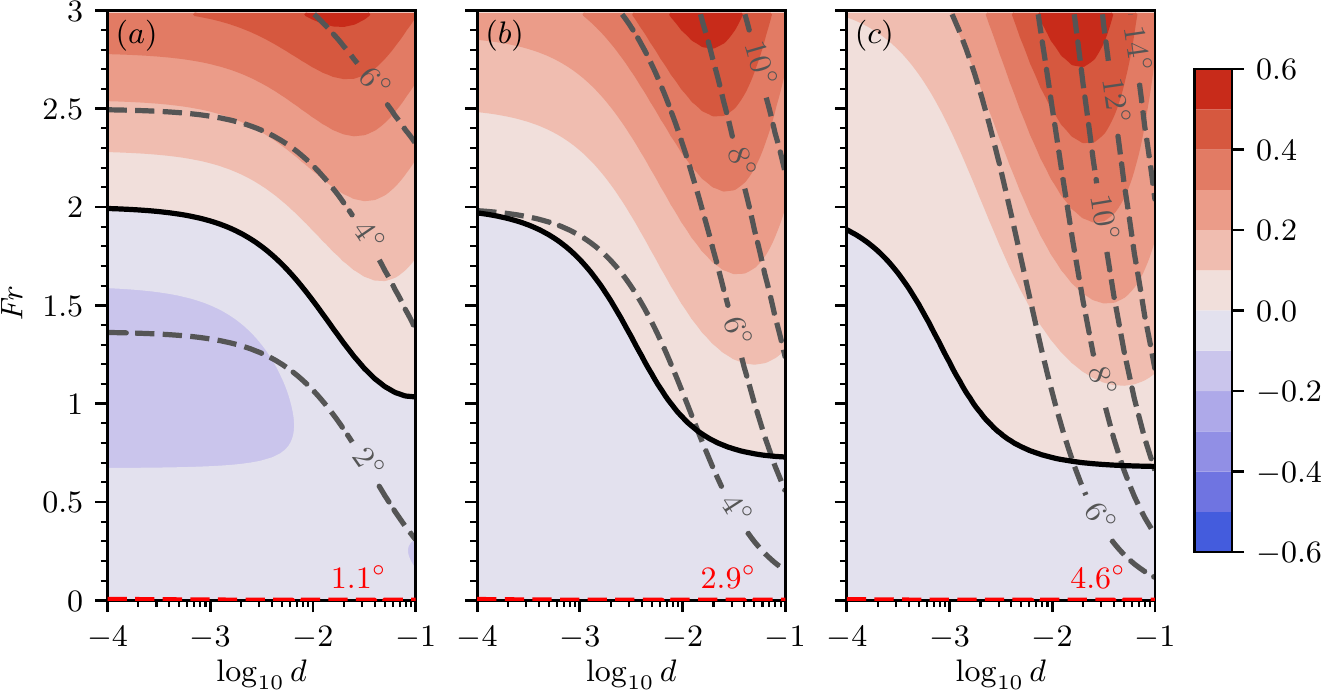}%
 \caption{%
     Existence of steady layers for our mixed drag formulation~\eqref{eq:mixed
     drag}, without morphodynamics.
     Dashed lines show the existence of steady flows
     at fixed slope angles as labelled, with the lowest red dashed line indicating
     the minimum slope angle $\phi = \arctan(\psi_0\mu_1)$.      %
     Filled contours show maximum values of the non-dimensional linear growth
     rates given in~\eqref{eq:hydraulic sigma}.
     Stable (blue) and unstable (red) regions are separated by the neutral
     stability boundary~\eqref{eq:crit Fr eqn} plotted in solid black.
     Successive panes represent increasing solid fractions $\psi_0$ from left to
     right as follows: ($a$)~0.2, ($b$)~0.5 and ($c$)~0.8.
     %
 } 
 \label{fig:growth rates}
\end{figure}
Dashed contours trace out the unidimensional family of steady layers for each
slope angle, across $(d, \Fr)$-parameter space, computed from~\eqref{eq:steady
drag phenom}, with the three panes corresponding to different fixed solid
fractions, $\psi_0 = 0.2$, $0.5$ and $0.8$, from left to right.  The minimum
slope angles for steady flows (dashed red contours) follow the line $\Fr = 0$.
Red and blue filled contours show the asymptotic growing and decaying growth
rates of perturbations respectively, computed by substituting $\tau_{u_0}$ and
$\tau_{h_0}$ from above into the limiting formula given earlier
in~\eqref{eq:resig lim}.  These are separated by the neutral stability
boundary~\eqref{eq:crit Fr eqn}, which is displayed in solid black.  When $d \ll
1$ (small particles, relative to the flow depth), the drag is dominated by the
Ch\'ezy term. In this regime, which covers most physically realisable grain
sizes, growth rates are essentially independent of $d$ and the stability
boundary is $\Fr \approx 2$.  Increasing $d$ outside this region leads to less
stable flows and lowers the stability boundary.  Increasing the solid fraction
generally leads to less severe growth rates, but decreases the range of slope
angles that permit stable steady flows (through increasing the minimum slope
angle).
We find the qualitative properties of this plot to be largely insensitive to our
specific choices of $C_d$, $\mu_1$, $\mu_2$ and $\beta$ whose values were given
in table~\ref{tab:illustrative params}.

When morphodynamics is non-negligible, steady flows must also
satisfy~(\ref{eq:steady morpho balance}\emph{b}).  That is, erosion and
deposition must be everywhere in balance,~$E_p(h,u,\psi) = D_p(\psi)$.  This
condition dictates the solid fractions where mass transfer can be in
equilibrium. Despite our efforts to obtain simple closures
in~\eqref{eq:erosion phenom} and~\eqref{eq:deposition phenom}, it is a complex
nonlinear equation that depends on many physical parameters.
Nevertheless, we can determine some generic properties of solutions.

\begin{figure}
    \begin{centering}
    \includegraphics{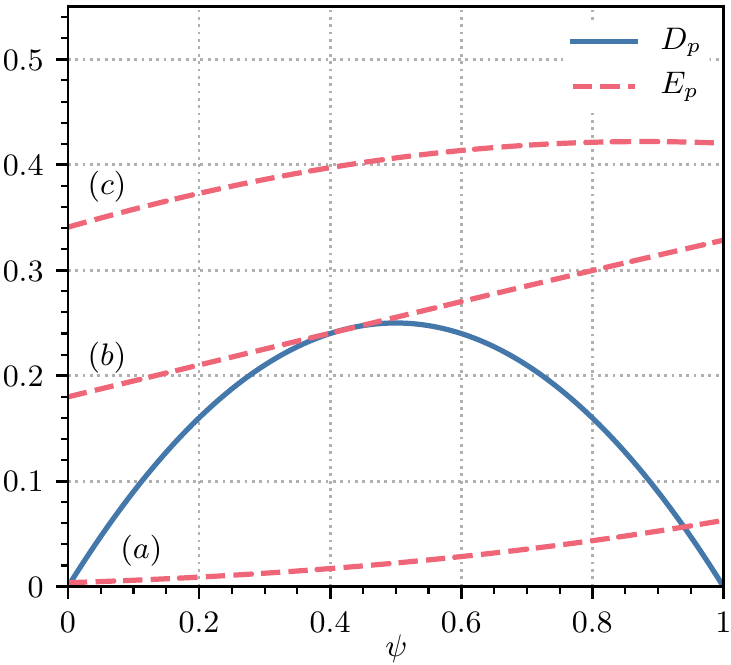}%
    \caption{Example deposition and erosion closures, as functions of the solid
        fraction $\psi$.
    Steady flows occur where $D_p$ and $E_p$ intersect.
    There are three cases:
    ($a$)~two steady states, one dilute and one concentrated ($\Fr = 0.75$);
    ($b$)~a single steady state where $E_p$ is tangent to $D_p$ ($\Fr \approx 2.63$);
    ($c$)~erosion always exceeds deposition (no steady states) ($\Fr = 3.25$).
    %
}
    \label{fig:erodep}
    \end{centering}
\end{figure}

We first consider the system parameters to be fixed (but arbitrary) and suppose
that the flow is in uniform steady balance with its drag (so $h = u = 1$),
leaving $D_p$ and $E_p$ functions of $\psi$ only. We also assume that $\theta >
\theta_c$, so that there is some particle entrainment. Then, in particular, 
$E_p(0) > 0$ and $E_p(1) > 0$.
Conversely, the deposition rate curve~\eqref{eq:deposition phenom} always obeys
$D_p(0) = D_p(1) = 0$. 
Therefore, since $E_p - D_p > 0$ for both $\psi = 0$ and~$1$, either: 
(\emph{a})~there exist an even number of coexistent
steady flows with different solid fractions,
(\emph{b})~erosion balances deposition
exactly at a turning point in $E_p-D_p$ or 
(\emph{c})~erosion always exceeds deposition.
We visualise these three cases in figure~\ref{fig:erodep}, where we plot $D_p(\psi)$
and $E_p(\psi)$ at different values of $\Fr$, using the illustrative model
parameters of table~\ref{tab:illustrative params}.
Aside from where explicitly stated otherwise, these parameters are fixed for the
remainder of this section.  Note that our choice of $D_p$ does not depend on
$\Fr$, while the Froude number dependence of $E_p$ enters through the basal drag
term in the Shields number.

Case ($a$), where there are multiple steady flows, occurs at lower $\Fr$
numbers. Here, erosion increases with solids concentration, intersecting the
deposition curve at two points. This leads to two corresponding steady flows:
one dilute and one relatively concentrated. Additional intersections (leading to
three or more steady flows) are a possibility, but would require a very
particular erosion curve. Physical intuition suggests that the dilute solution
is stable, since perturbations in either direction cause negative feedback:
decreasing $\psi$ away from this state leads to $E_p > D_p$, while increasing
$\psi$ leads to $E_p < D_p$.  Likewise, the concentrated solution (at $\psi
\approx 0.94$ in figure~\ref{fig:erodep}) invites either runaway deposition (if
$\psi$ decreases) or runaway erosion (if $\psi$ increases).  This process
suggests a possible mechanism underlying sediment distribution in debris flows,
which commonly feature an unsteady highly concentrated front trailed by a steady
dilute layer~\citep{Pierson1986,Hungr2000,Ancey2001}.
As $\Fr$ increases in figure~\ref{fig:erodep}, the pair of steady flows
coalesces at a single point; this is case~($b$). Beyond this point, no steady solutions
exist, case~($c$).  Here, erosion everywhere exceeds deposition. Uniform layers
in this case can never be truly steady, since they can only satisfy one
of~(\ref{eq:steady morpho balance}\emph{a,b}). If the drag is ever in
equilibrium with gravitational forces, then net entrainment injects
material into the flow.

\subsection{Global morphodynamic modes}
\label{sec:global morpho}%
The instability mechanism identified in figure~\ref{fig:erodep} is purely
morphodynamic and depends on a straightforward criterion: states are unstable
to this mode if $\Gamma_{\psi_0} > 0$. The process is
fundamentally an instability to uniform perturbations in flow concentration,
though since there is no intrinsic dependence upon spatial gradients we might
anticipate that it manifests as a destabilising feature at all wavenumbers.
However, this picture is a simplification, since it omits feedbacks from the
other flow fields. The full situation for uniform disturbances is contained
within our analysis of zero-wavenumber perturbations in \S\ref{sec:k = 0}, where
we computed the general growth rates of the four different linear stability
modes for $k=0$, two of which are always neutrally stable. As discussed earlier, if the
approximation of small $\Gamma_{u_0}$ can be made, the two remaining modes may
be understood simply: one is inherited from the hydrodynamic stability problem,
the other contains morphodynamic feedbacks.  Their growth rates are given
in~(\ref{eq:approx k = 0 sigma}\emph{a,b}).  The morphodynamic rate in
~(\ref{eq:approx k = 0 sigma}\emph{b}) may by understood as a competition
between two mechanisms for growth in $h$ and $\psi$. The process for~$\psi$
(when $\Gamma_{\psi_0} > 0$) has already been outlined. If $\Gamma_{h_0} > 0$,
then a small increase in the steady layer height leads to net entrainment which,
via~(\ref{eq:sw morpho nondim}\emph{a}), enhances growth in $h$ in turn.
Likewise, a small decrease in $h$ would cause the depth to decrease away from
its steady value. Conversely, if $\Gamma_{h_0} < 0$, then this term is
stabilising.
If $\Gamma_{h_0}$ is negligible with respect to $\Gamma_{\psi_0}(1-\psi_0)$,
then the morphodynamic mode has approximate growth rate $\sigma^*_m$, defined by 
\begin{equation}
\sigma_m^* = \Gamma_{\psi_0}(1- \psi_0).
\label{eq:sigma m star}
\end{equation}
In this case, stability is only governed by the mechanism for concentration
growth encapsulated by figure~\ref{fig:erodep}.

The accuracy of the above physical picture depends on the reliability of the
approximations made in reaching~(\ref{eq:approx k = 0 sigma}\emph{a,b})
and~\eqref{eq:sigma m star}.
These estimates are plausible (especially at lower $\psi_0$), since we might
expect the relative steepness of the hindered settling curve (plotted in
figure~\ref{fig:erodep}) to be more significant than gradients in $E_p$, which
are solely responsible for dictating $\Gamma_{h_0}$, $\Gamma_{u_0}$ and
depend on the small parameters $\varepsilon$ and $C_d$.

\begin{figure}
\centering%
\includegraphics{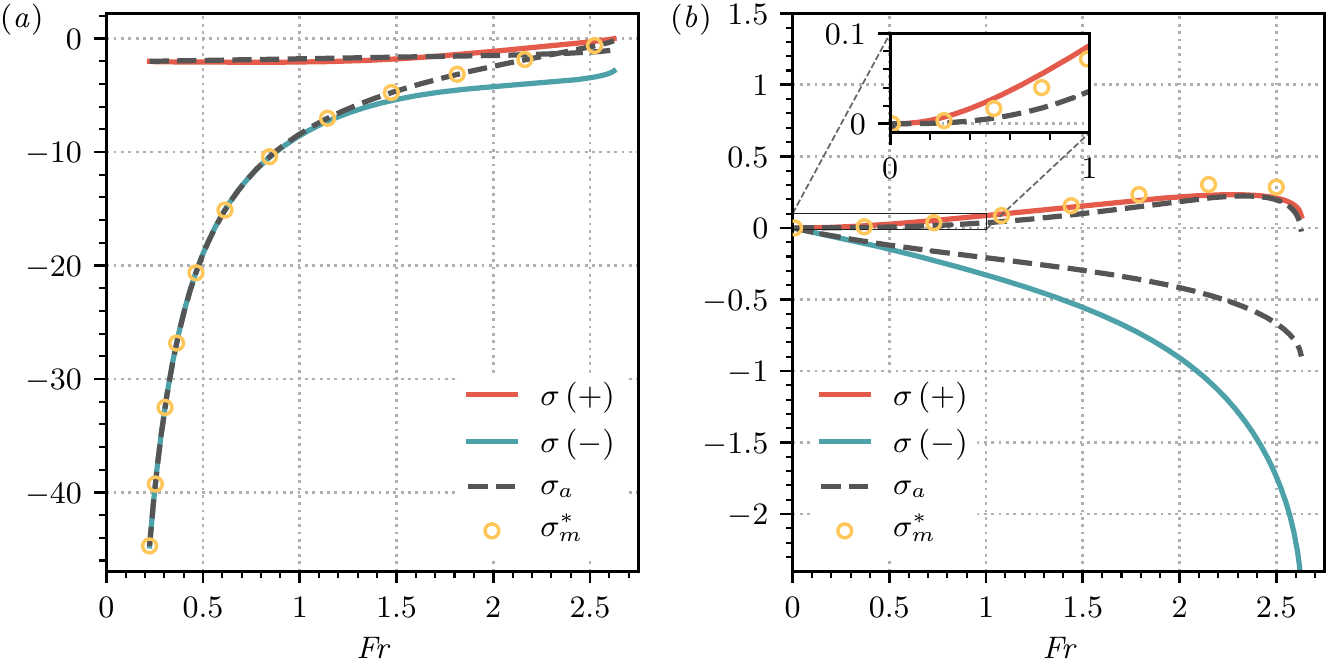}%
\caption{%
Growth rates for uniform ($k=0$) perturbations as a function of $\Fr$, for the
(\emph{a}) dilute and (\emph{b}) concentrated steady solution families
identified in figure~\ref{fig:erodep}. Solid curves show the two non-zero branches
of the exact growth rate $\sigma$,
computed from the formulae in~(\ref{eq:k = 0
sigma}\emph{b})--\eqref{eq:sc}.  The signs given in the legend indicate the
branch, according to the sign of $\pm\sqrt{s_c}$ in~(\ref{eq:k = 0
sigma}\emph{b}).
Also plotted are two approximations to the exact rates, $\sigma_a$ (dashed
lines), as defined in~\eqref{eq:approx k = 0 sigma}, and $\sigma_m^*$ (yellow
circles), as defined in~\eqref{eq:sigma m star}.
}
\label{fig:gamma derivs}
\end{figure}

For a more detailed analysis,
in figures~\ref{fig:gamma derivs}(\emph{a}) and
(\emph{b}) we show both non-zero branches of the exact growth rates
$\sigma$ (solid lines) for global disturbances (which are purely real), as given in~(\ref{eq:k = 0 sigma}\emph{a})--\eqref{eq:sc}, for
(\emph{a}) dilute and (\emph{b})~concentrated states.
On the same axes, we plot both approximations to the rates: the more general
formulae from~(\ref{eq:approx k = 0 sigma}\emph{a,b}), which we denote
$\sigma_a$ (dashed lines), and the cruder approximation to the morphodynamic
mode rate $\sigma_m^*$ (yellow circles), made above in~\eqref{eq:sigma m
star}.
For the dilute states, we confirm that $\sigma < 0$ across the range
where steady layers exist
($0.22 \lesssim \Fr \lesssim 2.62$). Moreover, both branches of $\sigma$ are well
approximated by $\sigma_a$ for $\Fr \lesssim 1.5$, and $\sigma_m^*$ lies very
close to its corresponding branch of $\sigma_a$, indicating that
$\Gamma_{\psi_0}$ is indeed primarily responsible for setting the sign of
$\sigma$ in this case. 
At higher $\Fr$, the approximating curves are less accurate.  However, this is
to be expected.
Whereas at low $\Fr$, the dilute solutions have $\psi_0 \approx 0$, which is where the
hindered settling curve is at its steepest (and consequently where
$|\Gamma_{\psi_0}|$ can be considered large compared with other derivatives
of~$\Gamma$), for higher $\Fr$ states have higher $\psi_0$
and the approximations of negligible $|\Gamma_{h_0}|$ and
$|\Gamma_{u_0}|$ gradually break down as $\psi_0$ approaches the turning point
in $D_p$ (see figure~\ref{fig:erodep}). However, we note that throughout,
$\sigma_m^*$ lies close to $\sigma_a$ since $|\Gamma_{h_0}|$ is small for dilute
states.  Furthermore, we need only be strictly concerned with $\Fr \lesssim 1$,
since outside this regime layers are susceptible to high wavenumber
instabilities.

By contrast, the approximations to the growth rates for the concentrated
solutions, in figure~\ref{fig:gamma derivs}(\emph{b}), are not especially good,
as highlighted by the figure insert. Therefore, $\Gamma_{u_0}$ cannot be
neglected here. Unfortunately, while perturbations in $h$ and $\psi$ and their
respective feedbacks may be understood in simple physical terms, this is not
easy to do in general for $u$, due to the many interacting contributions to
momentum present in the governing equations.  However, in this particular case,
we note that the intuition that concentrated steady states are typically
unstable is borne out, meaning that the feedbacks omitted in making the
approximation $\sigma_a$ are not stabilising on aggregate.  In fact, this is
guaranteed, since the pair of steady states in figure~\ref{fig:erodep} arises in
a saddle-node bifurcation as $\Fr$ is decreased from infinity. This implies
that, since the dilute flow is stable, the concentrated flow must have at least
one unstable direction.

\subsection{Linear growth rates for general wavenumbers}
Instability to uniform disturbances is not the only feature introduced by the
presence of morphodynamics, as our earlier analysis in \S\ref{sec:bed exchange}
indicates, since states are also vulnerable to the high-wavenumber
instability near unit Froude number.
Therefore, we broaden the discussion to incorporate the full linear
stability problem.  We begin by numerically solving the eigenproblem
in~\eqref{eq:morpho eigenproblem} using our chosen model closures, over a range
of finite wavenumbers.
Each of the four eigenmodes in the problem possesses a linear growth rate
continuously parametrised by~$k$.
As in~\S\ref{sec:k gg 1}, we label the modes I--IV according to the ordering of
their asymptotes used in~(\ref{eq:invisc growth rates}\emph{a--d}).  Recall that
in the high-$k$ regime, modes I \& II are analogues to the modes of the purely
hydraulic stability problem, whereas III \& IV are additional morphodynamic
modes that, involve perturbations in the solid fraction and bed
surface respectively.

\begin{figure}
\centering%
\includegraphics{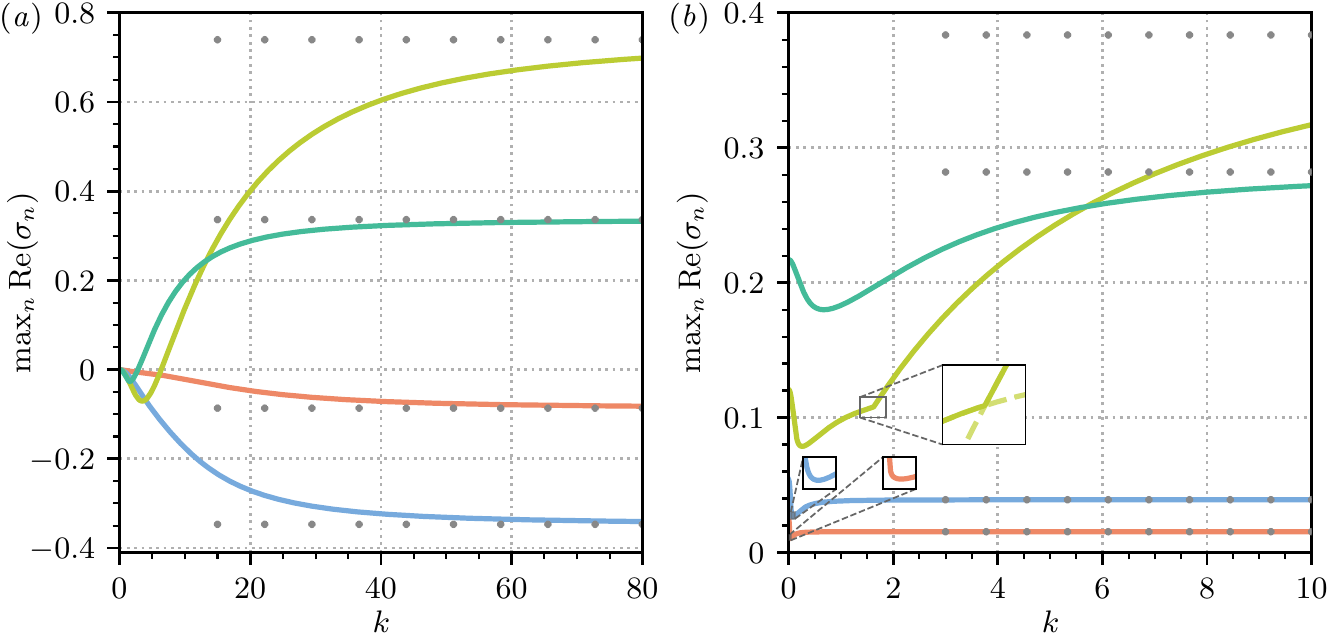}%
\caption{%
Perturbation growth rates for morphodynamic states as functions of wavenumber in
the unregularised problem~($\nu = 0$).  
The curves were computed by taking the maximum real
part of the four normal mode growth rates arising from~\eqref{eq:morpho
eigenproblem}, across a range of wavenumbers $k$, for states on the
(\emph{a})~dilute and (\emph{b})~concentrated solution branches
identified in figure~\ref{fig:erodep}.
The Froude numbers are $\Fr = 0.5$
(orange), $0.75$ (blue), $1.25$ (olive), $2$ (teal). High-$k$ asymptotes,
computed from maxima of the four expressions~(\ref{eq:invisc growth rates}\emph{a--d}), are plotted in dotted
grey. 
}%
\label{fig:growth rates no visc}%
\end{figure}

We denote the growth rates of each mode, indexed by wavenumber as $\sigma_n(k)$
for $n = 1,\ldots,4$. In figure~\ref{fig:growth rates no visc}(\emph{a}) we plot
$\max_n \Real(\sigma_{n})$ versus $k$, for states on the dilute solution branch.
Additionally, we plot their limiting high-$k$ growth rates, taken from the
maxima of the four expressions derived in~(\ref{eq:invisc growth rates}\emph{a--d}), in
dotted grey. 
Each curve passes through the origin, since the maximum growth at $k=0$ is given by
the neutral modes for these states.
The $\Fr = 0.5$ and $0.75$ solutions are stable to all perturbations, just as
they would be in the non-erosive case, with the latter solution being more
strongly damped.  The $\Fr = 1.25$ curve is stable to long-wavelength
disturbances and becomes unstable at $k \approx 6$. Its maximum over all $k$ is
given by its asymptotic value, to which it converges more slowly than the other
curves.  This solution would be stable in the non-erosive case: given the same
solid fraction ($\psi_0 \approx 0.02$), according to~\eqref{eq:crit Fr eqn},
non-erosive states turn unstable at $\Fr \approx 1.98$.  The $\Fr = 2$ state is
stable for a narrower range of wavelengths, becoming unstable at $k \approx 3$.
This state (which has $\psi_0 \approx 0.1$) would also be unstable in the
non-erosive situation. Its asymptotic growth rate is lower than the $\Fr =
1.25$ curve, which we shall see shortly is because it lies further from the
$\Fr = 1$ singularities present in~(\ref{eq:invisc growth rates}\emph{b,d}).
Aside from a narrow interval ($k \lesssim 1.5$) at small $k$ in the $\Fr = 2$
case where mode I dominates (not visible at this scale), each of these maximum
growth rates for dilute states are everywhere given by the growth of the
morphodynamic mode~IV.

In figure~\ref{fig:growth rates no visc}(\emph{b}), we plot the corresponding
maximum growth rate curves for the concentrated solution branch. These match the
intuition from~\S\ref{sec:global morpho}, that concentrated states should be
everywhere unstable (since we expect the basic global instability mechanism to
persist regardless of $k$). Growth at $k=0$ is a local maximum for each $\Fr$
and the corresponding instability is due to mode~III, which dominates over all
$k$ for $\Fr = 0.5$, $0.75$ and $\Fr = 2$. (The two lower $\Fr$ growth rate
curves turn sharply at $k \approx 0.025$ and $0.075$ respectively, but are
nevertheless smooth, as indicated on the insert.) Conversely, the $\Fr = 1.25$
curve is formed by a crossing of growth rates for modes~III and~IV, the latter
of which is neutral at $k = 0$.  The crossing point (at $k \approx 1.6$) is
shown in an insert, with the subdominant portions of the mode~III and~IV curves
also included (dashed lines). 

\begin{figure}
\centering%
\includegraphics{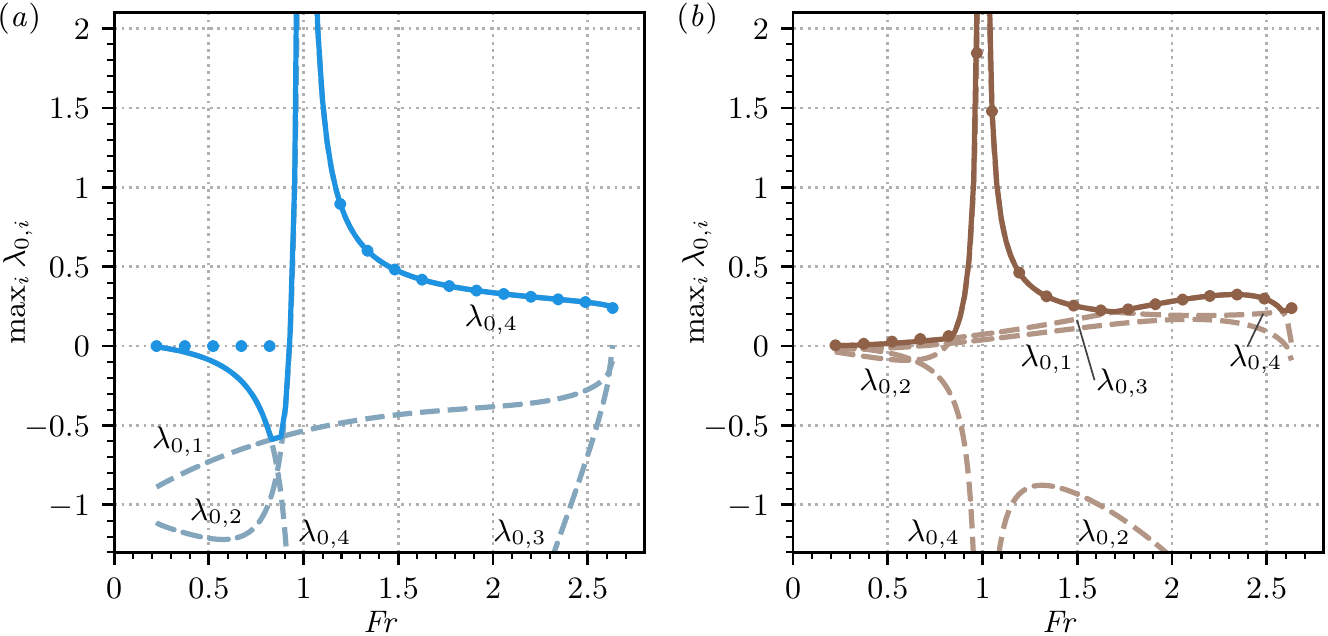}%
\caption{%
The maximum of the high-wavenumber asymptotic growth rates,
$\max_i\lambda_{0,i}$, plotted with solid curves as a function of~$\Fr$, for
the (\emph{a})~dilute and (\emph{b})~concentrated solution branches. With
dashed lines we plot the individual $\lambda_{0,i}(\Fr)$ curves, as labelled
for $i=1,\ldots,4$.  Also plotted (filled circles) are the maximum growth rates
of the corresponding solutions, over all wavenumbers, i.e.\
$\max_{k,n}\Real[\sigma_n(k)]$.
}
\label{fig:max growth vs Fr}
\end{figure}

The variations of the modes as functions of Froude number are encapsulated in
figure~\ref{fig:max growth vs Fr}.
Here, we plot the asymptotic ($k \gg 1$) growth rates $\lambda_{0,i}$, given
in~(\ref{eq:invisc growth rates}\emph{a--d}), for $i=1,\ldots 4$, with dashed lines. Their
maximum for each $\Fr$ is overlaid as a solid line. With filled circles, we plot the maximum
growth rate over all $k$.  Therefore, any discrepancy between the solid lines
and circles indicates that maximal growth is attained at some finite wavenumber.
Figure~\ref{fig:max growth vs Fr}(\emph{a}) presents the data for the dilute
solutions.
At low $\Fr$, solutions are stable and the maximum possible growth is zero, due to the
($k=0$) neutral modes. The asymptotic growth rate is dictated by the mode~IV curve, which
is briefly surpassed by mode~I at $\Fr \approx 0.85$, before growth is dominated
by the singular behaviour of modes~II (for $\Fr < 1$) and~IV (for $\Fr > 1$),
which diverge at $\Fr = 1$ with opposite sign. This induces instability
at $\Fr \approx 0.9$.
For all higher Froude numbers, the solutions remain unstable, even though they
would be sufficiently dilute to remain stable well past $\Fr = 1$ if
morphodynamic effects were neglected.
We also note that the asymptotic rate correctly identifies the onset of
instability and matches the maximum rate thereafter, thereby justifying the
focus on short wavelengths in our analysis.

The corresponding curves for the concentrated solution branch are shown in
figure~\ref{fig:max growth vs Fr}(\emph{b}).  As expected, this also features a
singularity at $\Fr = 1$ and is everywhere unstable.  The maximum growth rates
(filled circles) exactly match the corresponding asymptotic limits, except at
low Froude number ($\Fr \lesssim 0.85$), where growth at $k=0$ is slightly
larger than in the $k\gg 1$ regime, as we saw in figure~\ref{fig:growth rates no
visc}(\emph{b}).  Mode~III is always unstable and dominates the other modes over
a large region.  Near the singularity it is surpassed by modes~II (for $\Fr <
1$) and~IV (for $\Fr > 1$) and it is briefly surpassed again by mode~IV near
$\Fr \gtrsim 2.6$, where the dilute and concentrated solution branches coalesce.

\subsection{Effect of bed erodibility}
\label{sec:effect of eps}%
In addition to the singularity introduced at unit Froude number, a further
effect of the morphodynamics on dilute states may by identified in
figure~\ref{fig:max growth vs Fr}(\emph{a}).  In the limit of weak sediment
entrainment, $\varepsilon\to 0$, mode~I must turn unstable at $\Fr = 2$, due to
the classical roll wave instability for Ch\'ezy bottom
drag~\citep{Jeffreys1925}. Conversely, in the regime of figure~\ref{fig:max
growth vs Fr} ($\varepsilon=0.01$), this mode remains stable throughout the
range of $\Fr$ where steady states exist ($0.22\lesssim \Fr \lesssim 2.62$).
Unstable growth only occurs for mode~IV.

We observe the effect of increasing $\varepsilon$ from a weakly erodible regime
by reproducing the figure~\ref{fig:max growth vs Fr}(\emph{a}) plot for dilute
states at various values of $\varepsilon$, in figure~\ref{fig:roll wave
alterations}.
\begin{figure}
\centering%
\includegraphics{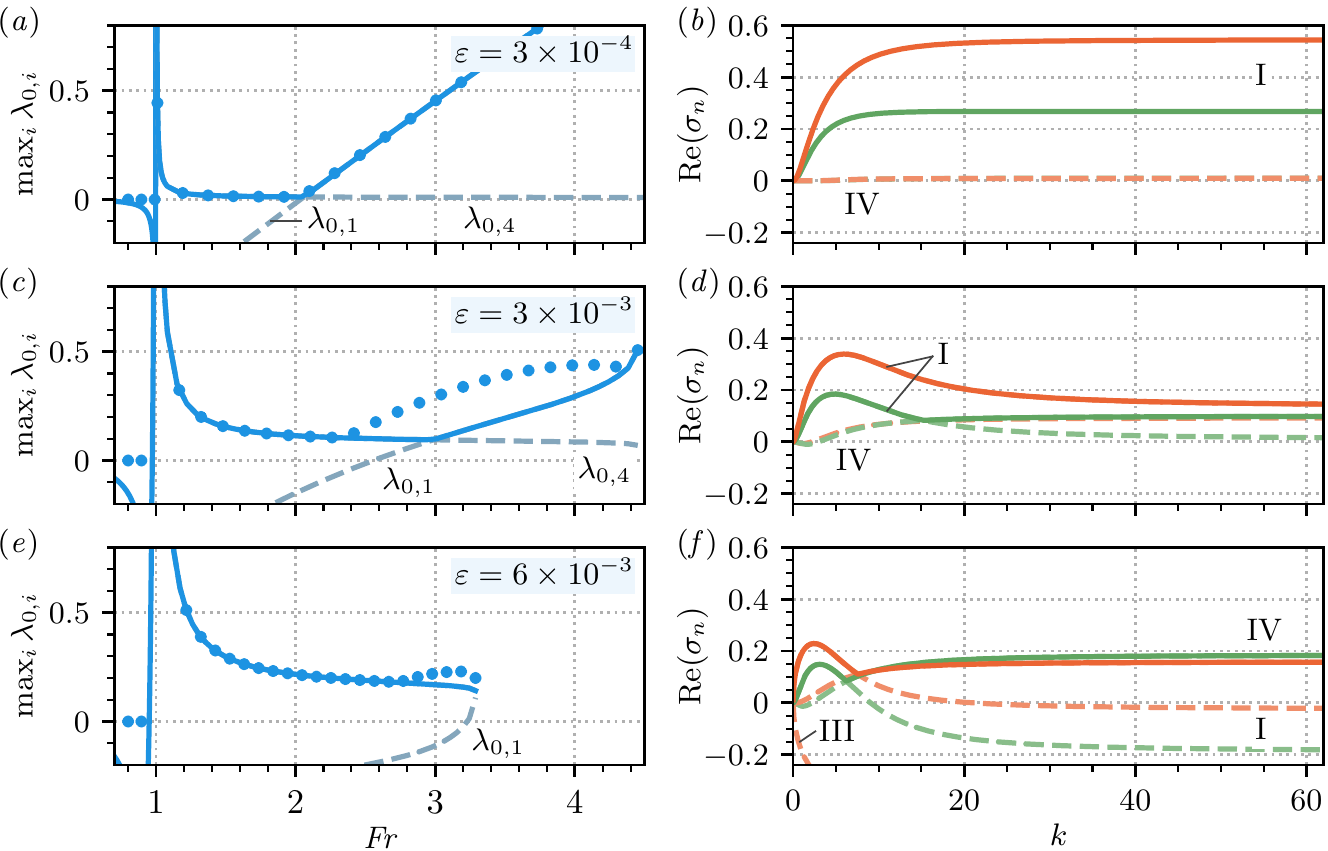}%
\caption{%
    Effect of the bed erodibility $\varepsilon$ on the growth rates of
    the unstable modes, showing the suppression of mode~I (associated with
    roll wave instability) as $\varepsilon$ increases.
    The left-hand plots~(\emph{a,c,e}) show maximum asymptotic growth rates
    ($\max_i \lambda_{0,i}$) as a function of $\Fr$ for dilute
    states (solid blue) and individual $\lambda_{0,i}(\Fr)$ curves (grey
    dashed), as labelled. Maximum growth rates over all wave numbers are 
    plotted with blue circles (as in figure~\ref{fig:max growth vs Fr}, which
    shows the $\varepsilon = 0.01$ case).  The right-hand plots~(\emph{b,d,f})
    show the growth rates $\sigma_n(k)$ of individual modes in dashed lines, as
    labelled, for $\Fr = 2.6$~(green) and $\Fr = 3.2$~(orange). The maxima of these
    curves are overlaid as solid lines.  The erodibility values for each row
    are
    (\emph{a,b})~$\varepsilon = 3\times 10^{-4}$,
    (\emph{c,d})~$\varepsilon = 3\times 10^{-3}$ and
    (\emph{e,f})~$\varepsilon = 6\times 10^{-3}$.
    }
\label{fig:roll wave alterations}%
\end{figure}
The other model parameters remain as stated in table~\ref{tab:illustrative
params}.
Beginning with the case of small $\varepsilon = 3\times 10^{-4}$, in
figure~\ref{fig:roll wave alterations}(\emph{a}), the signature of hydraulic
roll wave instability is clear. Mode I becomes unstable at a Froude number just
above $2$ and immediately dominates over the morphodynamic mode.  Outside the
singular point $\Fr = 1$, the high-$k$ growth rate of the latter mode (IV) is
$O(\varepsilon)$, which may be confirmed by consulting
the formula derived in~(\ref{eq:invisc growth rates}\emph{d}) and the closure
for entrainment in~\eqref{eq:erosion phenom}. Consequently, there is a narrowing
of the effective influence of the singularity, relative to the picture in
figure~\ref{fig:max growth vs Fr}(\emph{a}), and away from this region, mode~IV
is only very weakly unstable.  In figure~\ref{fig:roll wave
alterations}(\emph{b}), we plot $\Real(\sigma_n)$ as a function of $k$ for
modes~I \&~IV at $\Fr = 2.6$ (green) and $3.2$ (orange). Both curves for mode~I
are close to the corresponding unstable growth rates in the limiting case
$\varepsilon \to 0$ with purely Ch\'ezy drag, which asymptote to exactly $0.3$
and $0.6$ respectively, by equation~\eqref{eq:resig lim}.

We increase $\varepsilon$ by an order of magnitude in figures~\ref{fig:roll wave
alterations}(\emph{c--f}). When $\varepsilon = 3\times 10^{-3}$, growth in the
$1 < \Fr \lesssim 2.3$ range is now dominated by the morphodynamic mode and
$\lambda_{0,1}$ has substantially decreased, remaining negative until $\Fr \approx
2.56$. However, $\Real(\sigma_1)$ no longer attains its maximum in the high-$k$
limit.  Indeed, when $\Fr \gtrsim 2.3$, the most significant growth comes from
mode I at finite $k$, as evidenced by figure~\ref{fig:roll wave
alterations}(\emph{c}). By comparing the $\Real(\sigma_1)$ curves in
figure~\ref{fig:roll wave alterations}(\emph{d}) with their counterparts in
figure~\ref{fig:roll wave alterations}(\emph{b}) we see that growth of mode I is
suppressed in the short-wavelength regime as $\varepsilon$ increases.
On proceeding to $\varepsilon = 6\times 10^{-3}$, these trends continue. The
plots in figures~\ref{fig:roll wave alterations}(\emph{e,f}) show that growth of
the morphodynamic mode dominates over the range $1 < \Fr \lesssim 2.8$, before
being surpassed by mode I growth at low $k$. Interestingly, mode I is now only
unstable across a limited band of wavelengths. Numerical inspection at
indicates that at finite $k$, away from their asymptotic limiting
expressions in~\eqref{eq:O(1) eigvecs}, each eigenmode contains components of
all the flow variables $h$, $u$, $\psi$ and $b$. Therefore, at high enough
$\varepsilon$, there are no `purely hydraulic' instabilities, including
classical roll waves. However, this does not prohibit the existence of roll
waves that are intrinsically coupled with the bed and concentration dynamics.

Due to the complexity of the analytic expression for $\lambda_{0,1}$, given
in~(\ref{eq:invisc growth rates}\emph{a}) and~\eqref{eq:f pm} it is difficult to
appreciate directly why mode~I ultimately becomes suppressed when morphodynamic
effects are significant. Instead, we can look at a simplified case, where the
drag function is purely fluid-like, $\tilde\tau = \tilde \tau_f = C_d \tilde
\rho \tilde u^2$, and the bottom friction velocity $\tilde u_b$ is neglected.
Then, $\tau_{h_0} = \Gamma_{h_0} = \upsilon_0 = 0$, $\tau_{u_0} = 2$
and~(\ref{eq:invisc growth rates}\emph{a}) simplifies to
\begin{equation}
    \lambda_{0,1} = \frac{\Fr - 2}{2}
    + \frac{\Gamma_{u_0}}{4\Fr(\Fr + 1)}
    \left[
        (1 - \rho_b)(\Fr - 1) - 2\rho_b\Fr
    \right].
\end{equation}
The first term in this expression stems from the hydraulic limit and yields the
correct stability threshold in that case ($\Fr = 2$). Provided that both $\Fr >
1$ and $\rho_b > 1$ (\ie the bed density exceeds the bulk density), the second
term is strictly negative. Moreover, on
referring back to the mass transfer closures~\eqref{eq:erosion phenom}
and~\eqref{eq:deposition phenom}, we see that it is
proportional to $\varepsilon$.
Therefore, greater erodibility decreases $\lambda_{0,1}$ and correspondingly
increases the stability threshold for mode~I.

In the next two subsections, we investigate the effect of eddy viscosity and bed
load on linear growth rates in the morphodynamically dominated regime. We do not
independently investigate varying~$\varepsilon$ in these cases.  However, our
numerical observations indicate that, away from $\Fr = 1$, its essential effect
is preserved.  Namely, that as $\varepsilon$ increases from a negligible value,
the $O(\varepsilon)$ bed mode (IV) growth rates become larger and there is a
corresponding suppression of mode~I.

\subsection{Effect of eddy viscosity}
\label{sec:effect of regularisation}%
\begin{figure}
\centering%
\includegraphics{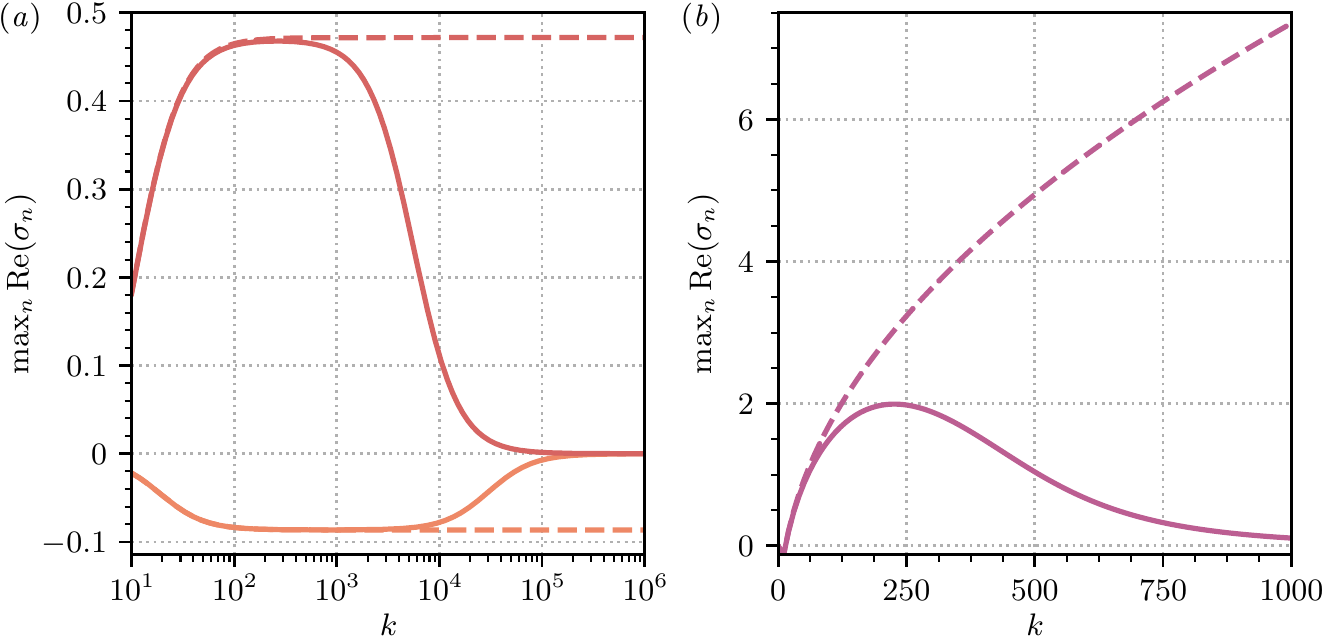}%
\caption{%
    Effect of the eddy viscosity regularisation term on maximum growth rates.
    Curves for the
    dilute solution branch are shown with $\nu = 0$ (dashed) and $10^{-4}$
    (solid). 
    The Froude numbers are:
    (\emph{a})~$\Fr = 0.5$ (orange) and $1.5$ (red); 
    (\emph{b})~$\Fr = 1$ (purple). 
}
\label{fig:growth rates visc}
\end{figure}

Figure~\ref{fig:growth rates visc} demonstrates the effect of the eddy viscosity
term studied in \S\ref{sec:regularisation}.
The size of the parameter $\nu$ sets the scale over which diffusive effects are
important. If chosen sufficiently small, the term only influences the high-$k$
regime. In figure~\ref{fig:growth rates visc}(\emph{a}), we plot maximum normal
mode growth rates for dilute solutions with $\Fr = 0.5, 1.5$, $\varepsilon =
0.01$, $\nu = 10^{-4}$ (solid lines) and compare these with the corresponding
rates for the unregularised system, $\nu = 0$ (dashed lines). At low $k$
(including $0 \leq k < 10$, not shown), the growth rates are essentially
unchanged by the presence of the dissipative term.  Then, at higher $k$, both
rates for the regularised problem converge to exactly zero, where they remain in
the high-$k$ limit. These may be compared with the analytical formulae
in~\eqref{eq:visc growth rates}. Computing them for our particular parameters
confirms that both cases are dominated by the asymptote $\sigma = 0 +
O(k^{-1})$, corresponding to perturbations of the bed~(\ref{eq:visc growth rates
all}\emph{b}). For the $\Fr = 0.5$ case, the growth rate increases when it
approaches this limit at high $k$.  However, note that since it approaches zero
from below, this does not affect flow stability. Therefore, we conclude that,
away from the singularity at $\Fr = 1$, the addition of the diffusive term
succeeds in damping out short-wavelength disturbances without affecting the
system outside the asymptotic regime.
The behaviour at $\Fr = 1$ is shown in figure~\ref{fig:growth rates
visc}(\emph{b}) and confirms the successful regularisation of the growth rate
singularity.  While the unregularised rate diverges like $\sim k^{1/2}$ (as
shown in~\S\ref{sec:k gg 1}), when $\nu = 10^{-4}$ it decays to zero within $0
\leq k \lesssim 10^3$. It reaches a maximum growth rate of approximately~$2$,
around 3--6 times the magnitude of unregularised growth rates either side of the
singularity at $\Fr = 0.75$ and $1.25$, plotted in figure~\ref{fig:growth rates
no visc}(\emph{a}). 

\begin{figure}
\centering%
\includegraphics{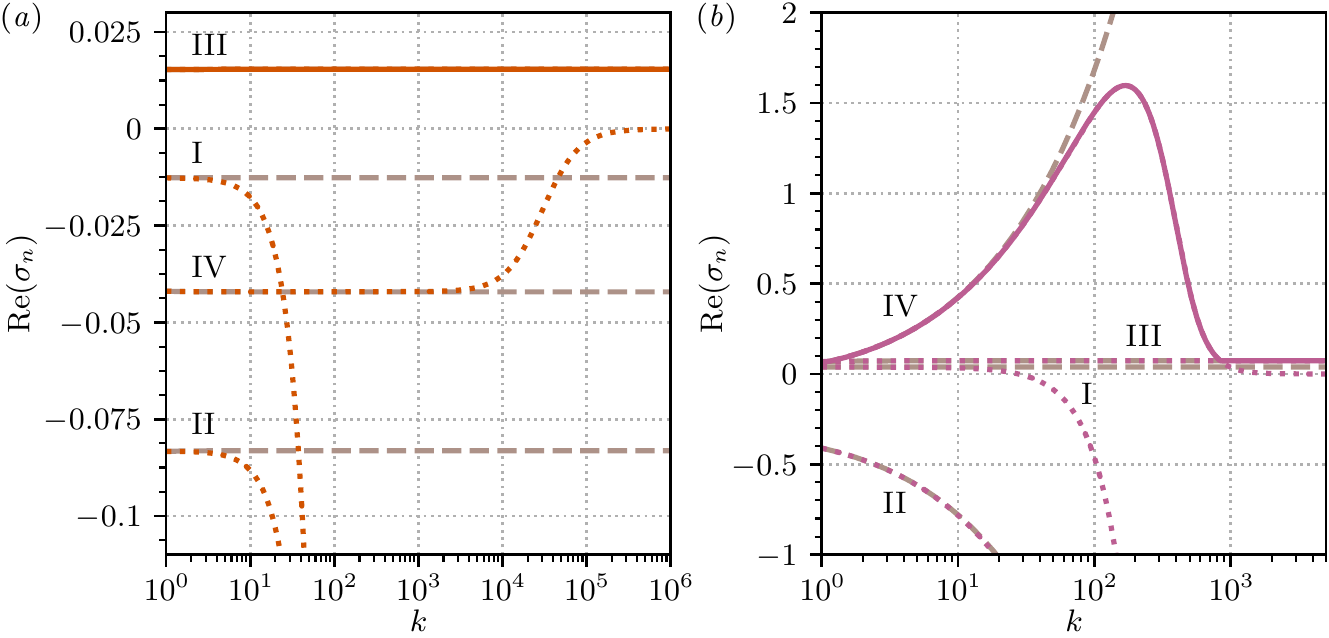}%
\caption{%
    Individual growth rates as a function of $k$ for concentrated steady states.
    The curves are labelled I--IV according to their corresponding linear
    stability modes in the $\nu = 0$ analysis of \S\ref{sec:k gg 1}. Unregularised rates are plotted with dashed grey lines;
    regularised rates ($\nu = 10^{-4}$) with dotted lines. The maximum 
    over all curves in the $\nu = 10^{-4}$ case is overlaid as a solid line.  The
    Froude numbers are (\emph{a}) 0.5 and ($\emph{b}$) 1.
}
\label{fig:growth rates visc conc}
\end{figure}

Figure~\ref{fig:growth rates visc conc} shows the effect of regularisation
on concentrated states. 
In these plots we include additional detail, plotting the growth rates for all
four modes.  Dashed lines show curves with $\nu = 0$; dotted lines show the
corresponding rates with $\nu = 10^{-4}$.  We label these~I--IV by numerically
computing the asymptotic rates~(\ref{eq:invisc growth rates}\emph{a--d})
and~\eqref{eq:lambda half} for the unregularised system, which match the
corresponding regularised rates at lower~$k$.  The case of $\Fr = 0.5$, away
from the singularity is given in figure~\ref{fig:growth rates visc
conc}(\emph{a}).  Both hydrodynamic modes (I and II) are severely damped at
high~$k$. On checking these curves against the asymptotic rates
in~\eqref{eq:visc growth rates}, we find that mode I corresponds
to~(\ref{eq:visc growth rates all}\emph{a}), scaling as $\sim-k^2$
asymptotically, while mode II eventually converges to $\lambda_- \approx -4
\times 10^4 = -1/\nu\Fr^2$ in this case (outside the range of the figure axes).
The mode~IV rate increases as $k$ increases, eventually asymptoting to $0$.
Finally, as argued in \S\ref{sec:regularisation}, mode III, which asymptotes to
$\lambda_+$ ($\approx 0.015$), is not much affected by the eddy viscosity term,
provided that $\nu$ is small relative to the other terms in~\eqref{eq:lambda
pm}.  The situation at the $\Fr = 1$ singularity for concentrated states is
shown in figure~\ref{fig:growth rates visc conc}(\emph{b}). When $\nu = 0$, the
mode~II and~IV growth rates can be seen diverging to $-\infty$ and $+\infty$
respectively as $k\to\infty$. As established in \S\ref{sec:k gg 1}, these scale
like~$\sim\pm k^{1/2}$, asymptotically -- while modes~I and~III converge to
$\lambda_{0,1}$ ($\approx 0.04$) and $\lambda_{0,3}$ ($\approx 0.07$)
respectively. With the addition of regularisation, the modes qualitatively
mirror the $\Fr = 0.5$ case: the hydraulic modes are strongly damped, while
mode~IV no longer diverges and decays to zero, being eventually surpassed by
mode~III which is essentially unaffected by the eddy viscosity.

We have briefly experimented with introducing an additional diffusive term to the solid mass
transport equation. This takes the form $\frac{\partial~}{\partial
x}(\kappa h \frac{\partial \psi}{\partial x})$ and is added to the right-hand
side of~\eqref{eq:sw morpho nondim 2}, thereby contributing an additional
non-zero term $D_{23} = \kappa$ to the matrix $\matr{D}$ in the linear stability
eigenproblem~\eqref{eq:morpho eigenproblem visc}. The free parameter $\kappa =
\tilde \kappa / (\tilde u_0 \tilde l_0)$, where $\tilde \kappa$ is a (constant)
characteristic sediment diffusivity sometimes included in
models~\citep[e.g.][]{Balmforth2012,Bohorquez2015}.
Its inclusion is consistent with the basic physics of sediment transport.
However, the effect on the growth rates is modest, for the values explored
($10^{-3} < \kappa < 10^{-5}$), effectively serving only to
damp mode~III at high wavenumbers. Consequently, figure~\ref{fig:growth rates
visc} is unaffected, while the maximum unstable growth in figure~\ref{fig:growth
rates visc conc} ultimately rolls off at high $k$, with the roll-off being more
severe for larger values of $\kappa$. Therefore, in the context of simple
shallow flow modelling, sediment diffusion, combined with momentum diffusion via
eddy viscosity, could serve as an unobtrusive way to prevent
instabilities from developing over unphysically short length scales.

We now examine the effect of varying $\nu$. Figure~\ref{fig:max growth vs Fr
visc} shows the reduction of growth rate at the $\Fr = 1$ singularity as $\nu$
is increased over three orders of magnitude for both the (\emph{a})~dilute and
(\emph{b})~concentrated solution branches.
\begin{figure}
\centering%
\includegraphics{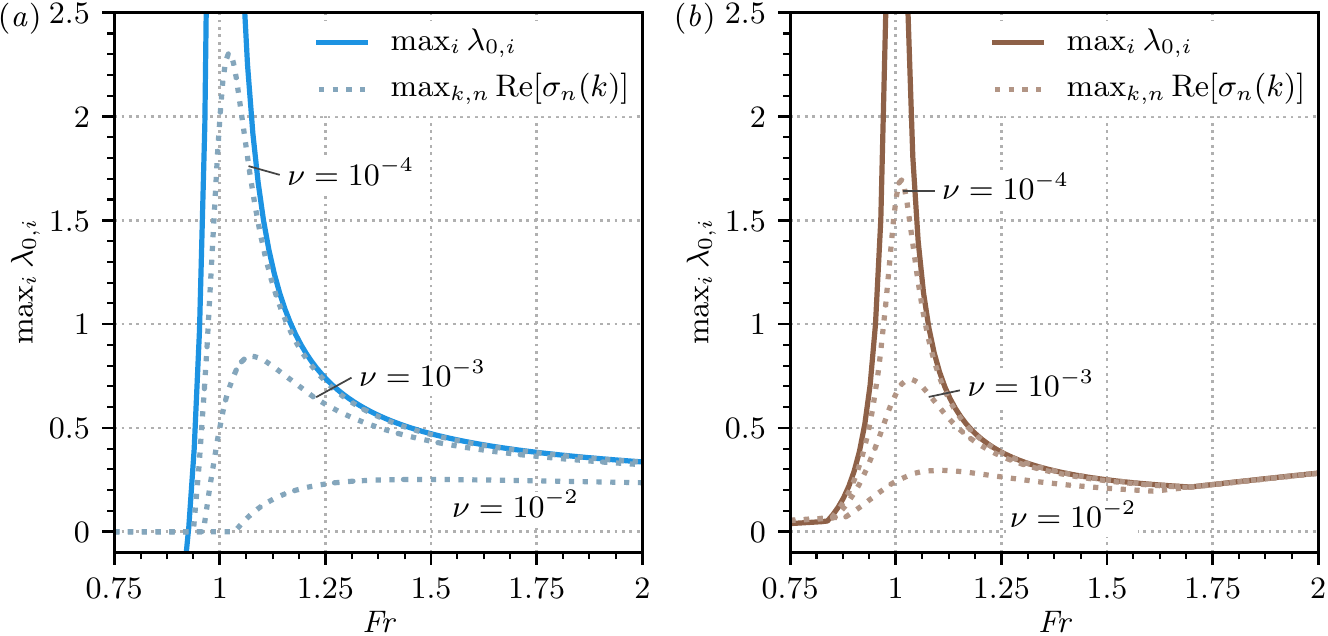}%
\caption{%
    Effect of eddy viscosity size on the severity of morphodynamic instability. Solid curves
    show the maximum asymptotic growth rates for the unregularised system (as
    shown previously in figure~\ref{fig:max growth vs Fr}), near the $\Fr = 1$
    singularity, for (\emph{a})~dilute and (\emph{b})~concentrated steady
    states. Dotted curves show the maximum growth rates over all wavenumbers
    and all four modes, for various $\nu$, as labelled.
}
\label{fig:max growth vs Fr visc}%
\end{figure}
At $\nu = 10^{-4}$, the maximum growth is curtailed only in a relatively small
neighbourhood of $\Fr = 1$. Consequently, growth still peaks sharply in this
region and elsewhere follows the asymptotic rate for the $\nu = 0$ case (solid
curves).  Increasing $\nu$ smooths over the signature of the singularity: its
(diminished) influence is clear at $\nu = 10^{-3}$, but by $\nu = 10^{-2}$ there
is only a residual trace of it. For the larger values of $\nu$, growth largely
fails to reach the asymptotic rates at all. We note also that increasing $\nu$
increases both the Froude number at which maximum growth occurs and, in the case
of dilute states, the onset of instability.

A constant eddy viscosity is a crude parametrisation of the effects of
turbulent dissipation.  Consequently, studies of non-erosive shallow layers have
typically treated selection of $\nu$ (when included) as a means to an end --
either to smooth over hydraulic jumps in flow profiles, or to constrain
instabilities to within a bounded
spectrum~\citep{Needham1984,Hwang1987,Balmforth2004}.  Provided $\nu$ is
sufficiently small, this is reasonable since roll wave onset and development
tends to be largely insensitive to the exact magnitude of the dissipation
term~\citep{Chang2000}.
However, figure~\ref{fig:max growth vs Fr visc} suggests this is not the case
when the evolution of the bed is accounted for.
Here, the size of $\nu$ necessarily dictates the severity of the morphodynamic
instability. A plausible range for $\nu$ can be ascertained as follows. Suppose
that dissipation acts over length and time scales set by shearing within the
turbulent boundary layer. Then we may anticipate $\tilde \nu \sim \tilde u_*
\tilde h$, where $\tilde u_*$ is the basal friction velocity, equivalent to the
closure used for $\tilde u_b$, given in~\eqref{eq:ub}.
For a dilute steady state flow, $\tilde u_* \approx \tilde u_0 \sqrt{C_d}$. In
our dimensionless units, $\nu = \tilde \nu g \sin\phi / \tilde u_0^3 = \tilde
\nu \tan\phi/ (\tilde h_0 \tilde u_0 \Fr^2)$ and (since $\tan\phi \approx C_d
\Fr^2$ in the dilute regime) therefore $\nu \sim C_d^{3/2}$. With our chosen
drag coefficient, this yields $\nu \sim 10^{-3}$. After accounting for potential
deviations from this rough order-of-magnitude estimate and especially the ways
in which suspended sediment may suppress turbulent fluctuations, it is difficult
to rule out any of the scenarios in figure~\ref{fig:max growth vs Fr visc} with
confidence.  However, it is at least reasonable to conclude that diffusive
effects are neither negligible, nor likely to completely quash unstable growth
near unit Froude number in morphodynamic shallow flow models.  

\subsection{Effect of bed load}
\label{sec:effect of bed load}%
In \S\ref{sec:bed load}, we showed that the inclusion of a bed load flux $Q$
also prevents ill posedness, provided that the constraint~\eqref{eq:hyperbolic
ineq} is satisfied. It does so by altering the system characteristics, thereby
avoiding a resonance between modes II \& IV at $\Fr = 1$ that otherwise leads to
unbounded linear growth.
In fluvial settings, the physical basis for such terms is well established
\citep[see e.g.][]{Gomez1991} and similar analyses to ours have noted the role
it plays in ensuring strict hyperbolicity of models
\citep{Cordier2011,Stecca2014,Chavarrias2018,Chavarrias2019}.  In more severely
concentrated situations, such as debris and purely granular flows, the case for
a bed load is less clear since it may not be possible to distinguish between
grains crawling along the bed surface and grains in suspension. Therefore, bed
load fluxes are not typically included in numerical models of these flows.
However, since bed load dictates the characteristic wave speed of the bed
surface, it may nonetheless have a role to play in these settings that is
not currently appreciated.

\begin{figure}
\centering%
\includegraphics{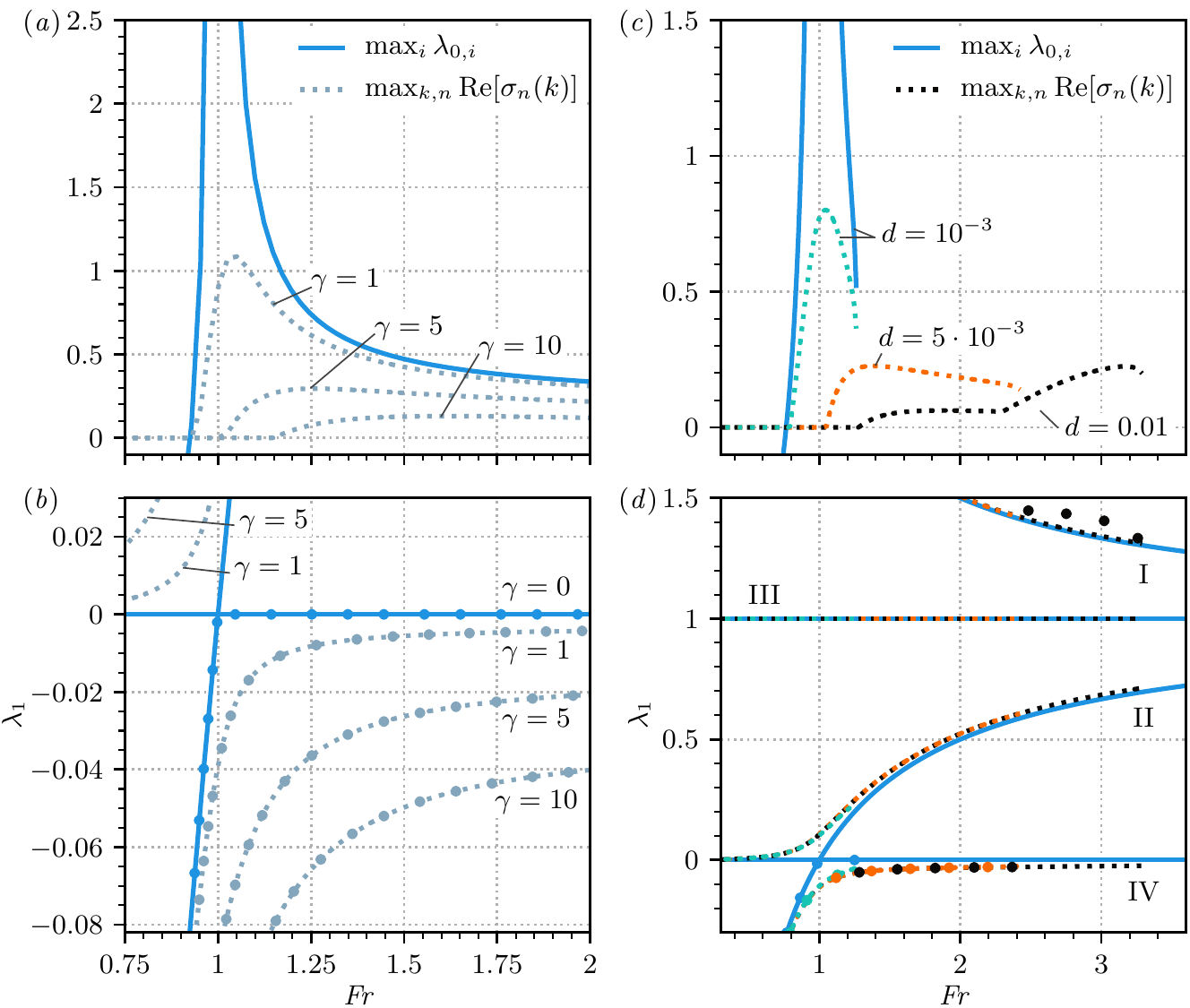}%
\caption{%
    Effect of a bed load flux term on linear growth rates and the corresponding
    instabilities. Part~(\emph{a}) shows curves of maximum growth rate over all
    modes and wavenumbers as a function of $\Fr$ (dotted), for $\varepsilon =
    0.01$, $d = 0.01$ and various $\gamma = 1$, $5$, and $10$ as labelled. 
    Also shown for reference, is the analytical curve of maximum growth rate at
    high wavenumber (solid blue), for $\gamma = 0$.
    Beneath this, in~(\emph{b}), we plot the system
    characteristics $\lambda_1(\Fr)$ with a solid blue line for $\gamma = 0$ and
    dotted lines for the $\gamma > 0$ cases. In the unstable regions, we also plot the
    corresponding wave speed
    $c_\mathrm{max}(\Fr)$ of the dominant unstable mode at its most unstable
    wavenumber (filled circles).
    Part~(\emph{c}) shows maximum growth rate curves, as in~(\emph{a}), for
    $\varepsilon = 6\times 10^{-3}$, $\gamma = 8$ and various $d = 10^{-3}$~(green dotted),
    $5\times 10^{-3}$~(orange dotted), $10^{-2}$~(black dotted).
    The reference curve $\max_i \lambda_{0,i}$~(solid blue) is plotted for
    $\gamma = 0$ and $d = 10^{-3}$.
    Underneath, in~(\emph{d}), the characteristics (solid and dotted lines for
    $\gamma = 0$ and $\gamma = 8$ respectively) and
    $c_\mathrm{max}(\Fr)$ (filled circles) are plotted using the same colour
    scheme as~(\emph{c}), with labels I--IV to indicate the corresponding
    eigenmodes.
    }
\label{fig:max growth vs Fr bedload}%
\end{figure}

We shall consider the effect of bed load separately from eddy viscosity
by resetting $\nu=0$ and employing the flux closure 
in~\eqref{eq:mpm}. 
Although the effect on concentrated steady states was briefly investigated, the
conclusions did not differ substantially from the relatively
dilute case, whose results we report below.
In figure~\ref{fig:max growth vs Fr bedload}(\emph{a}), we
plot the maximum growth rates over all modes and wavenumbers as a function of
$\Fr$ for various values of the bed load flux strength $\gamma$, increasing from
zero (solid blue) to $\gamma = 1$, $5$ and $10$ (dotted). 
The plot may be compared with figure~\ref{fig:max growth vs Fr visc}(\emph{a})
and shares the same essential features. This should not be surprising:
relatively small values of $\gamma$ do not displace the system characteristics
$\lambda_1$ far from their resonant values, while larger values push them far
apart, leaving little trace of the $\Fr = 1$ singularity.  We observe this
directly in figure~\ref{fig:max growth vs Fr bedload}(\emph{b}),
which plots $\lambda_1(\Fr)$ for each $\gamma$, as labelled (solid and dotted
lines), using the same horizontal axes. These curves are obtained by numerically solving~\eqref{eq:O(k) 1}.  The
solid lines are the characteristics $\lambda_1 = 1 - \Fr^{-1}$
(mode~II) and $\lambda_1 = 0$ (mode~IV) of the system without bed load, as derived
in~\eqref{eq:lambdas}.
As $\gamma$ increases from zero, the curves
diverge from the singular intersection point of these two
characteristics. 
Typical
values of $\gamma$ from the fluvial literature are $O(10)$~\citep[see
e.g.][]{Cordier2011}, around the upper end of the range considered here.  In
this regime, growth rates are relatively modest and the onset of instability
notably increases with $\gamma$, reaching $\Fr \approx 1.15$, when $\gamma =
10$.  Where solutions are unstable, we also plot, in figure~\ref{fig:max growth
vs Fr bedload}(\emph{b}), points along $c_\mathrm{max}(\Fr)$, which we define
here to be the wave speed $c = -\Imag(\sigma) / k$ of the dominant unstable mode
at its most unstable wavenumber, \ie the mode (for each $\gamma$) whose
corresponding growth rates are in figure~\ref{fig:max growth vs Fr
bedload}(\emph{a}). Each set of points lies exactly on its corresponding
characteristic curve.  This is because, in this case, the maximum growth at each
$\Fr$ occurs in the high-$k$ regime and $c\to\lambda_1$ as $k\to\infty$.
Furthermore, we note that since $c_\mathrm{max} < 0$ for $\gamma > 0$, dominant
perturbations travel upslope (and do so more rapidly if the bed load strength is
larger).  When $\gamma = 10$, the picture in figures~\ref{fig:max growth vs Fr
bedload}(\emph{a,b}) shares traits with some fluvial systems -- there is neither
ill posedness, nor severely accentuated growth near unit Froude number, and
the morphodynamics drives slowly upstream-migrating bedforms \citep[see
e.g.][]{Colombini2008,Seminara2010}. It is tempting to think of the
morphodynamic processes in this regime as being essentially `bed load dominant'.
However, pure bed load formulations do not feature instabilities near $\Fr =
1$~\citep{Lanzoni2006}. Indeed, we have checked that decreasing~$\varepsilon$ to
$10^{-4}$ (and retaining $\gamma = 10$), removes the morphodynamic instability
in our model, which then remains stable until the threshold for roll waves, near
$\Fr = 2$. Therefore, the bulk mass transfer term (\ie suspended load) plays a
role in sustaining the instabilities of figure~\ref{fig:max growth vs Fr
bedload}(\emph{a}).

Furthermore, even if $\gamma$ is large enough that the characteristics are well
separated, it is possible to see the influence of the $\Fr = 1$ singularity if
we move to a regime where suspended load is enhanced.  In figure~\ref{fig:max
growth vs Fr bedload}(\emph{c}), we fix $\gamma = 8$, $\varepsilon = 6 \times
10^{-3}$ and plot maximum growth rate curves, in the vein of figure~\ref{fig:max
growth vs Fr bedload}(\emph{a}), for different $d = 0.01$, $5\times 10^{-3}$ and
$10^{-3}$. Also shown for reference is the singular high-$k$ growth curve (solid
blue), for $d = 10^{-3}$ and no bed load.  
Note that these curves may only be plotted for $\Fr$ where steady flows exist
and this range shrinks as $d$ decreases. (This is because smaller particles are
more easily eroded, meaning that the unsteady scenario of
figure~\ref{fig:erodep}, where erosion always exceeds deposition, occurs at lower $\Fr$.)
The trend of figure~\ref{fig:max growth vs Fr bedload}(\emph{c}) shows that decreasing
particle size leads to more severe growth near $\Fr = 1$, with the $d =
10^{-3}$, $\gamma = 8$ case (dotted green) inheriting a severe instability in
this region. The corresponding characteristics, plotted in
figure~\ref{fig:max growth vs Fr bedload}(\emph{d}) below, are
not greatly affected by changes in $d$. Therefore, enhanced growth around unit
Froude number cannot be due to near intersection of characteristics in this
case. Without simple analytical expressions for the growth rates when $Q \neq
0$, it is difficult to pin down exactly why small $d$ has this effect. However,
we note that since our closures for erosion and bed load, given
in~\eqref{eq:erosion phenom} and~\eqref{eq:mpm} respectively, have similar
dependencies on the excess shear stress, it is straightforward to see for any
given $\Fr$, that $Q / E \propto d\gamma / \varepsilon$. Therefore, smaller $d$
implies that the magnitude of bed load is diminished, relative to the suspended
load dynamics.

Finally, just as we saw in figure~\ref{fig:max growth vs Fr bedload}(\emph{b}), we note that
for the $d = 10^{-3}$ and $d = 5\times 10^{-3}$ case, $c_\mathrm{max}$ (filled
circles) lies exactly on the characteristic curves corresponding to slowly
upstream-propagating disturbances.  However, for the $d = 0.01$ branch, this
mode is only dominant for Froude numbers up to approximately $2.4$, where the
rapidly downstream-propagating perturbations of mode~I become the most unstable.
This is marked in figure~\ref{fig:max growth vs Fr bedload}(\emph{c}) by a
steepening of the maximum growth rate curve and may be compared to the curves in
figures~\ref{fig:roll wave alterations}(\emph{a--d}), which depict an analogous transition between
modes~IV and~I in the $Q=0$ setting. The latter mode is related to hydraulic
roll wave instability, as discussed in~\S\ref{sec:effect of eps}. Note that in this region
($\Fr \gtrsim 2.4$), $c_\mathrm{max}$ does not precisely follow the mode~I
characteristic. This is because, as in figure~\ref{fig:roll wave alterations},
maximum growth occurs at finite~$k$, rather than in the asymptotic regime where
disturbance wave speeds are given exactly by the characteristics.

\subsection{Summary}
\begin{figure}
 \centering%
 \includegraphics{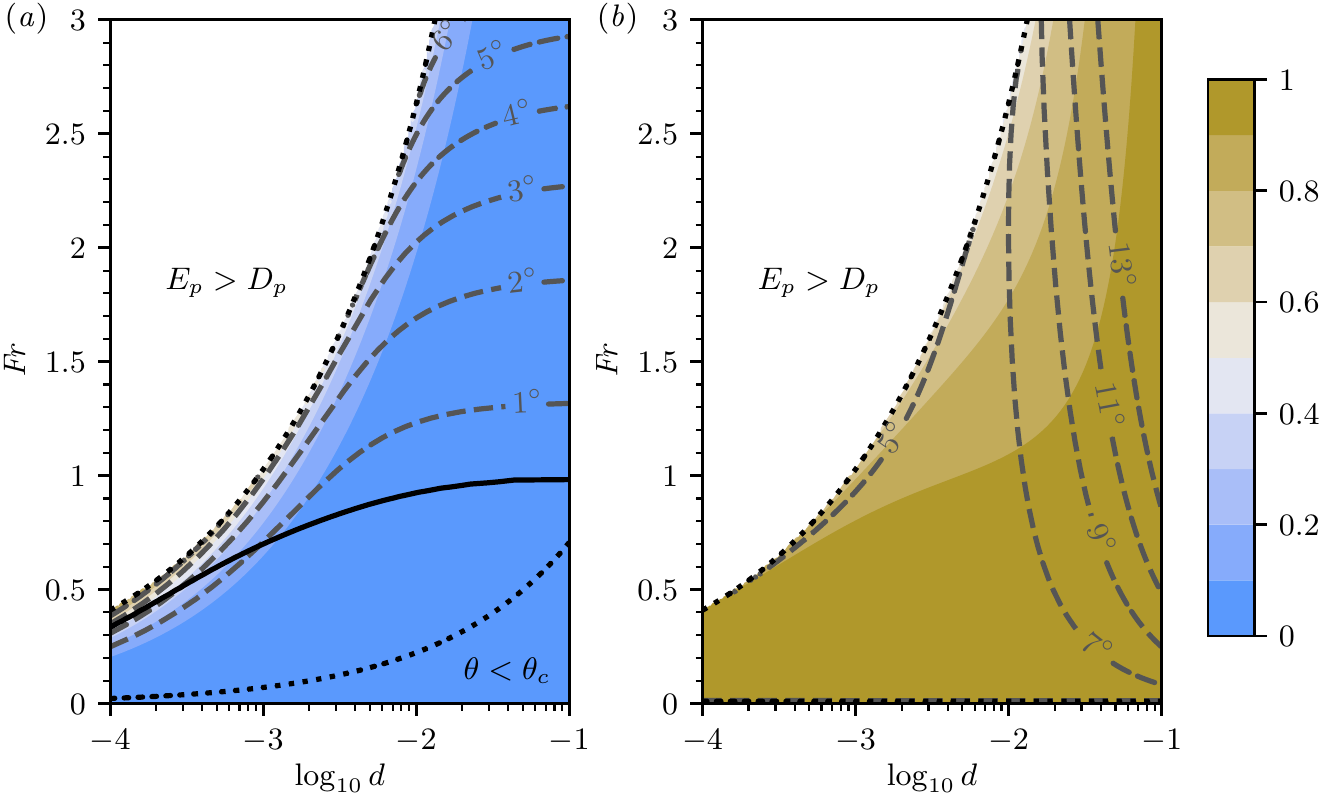}%
 \caption{%
     Existence and stability of uniform steady layers in the morphodynamic case.
     Plotted on two axes are filled contours of scaled solid fraction $\psi$ (shaded regions) for the
     (\emph{a})~dilute and (\emph{b})~concentrated solution
     branches, over a representative range of $d$ and $\Fr$ values. The region
     where steady erosive flows exist is outlined by dotted black lines.
     Where no steady flows exist, the plot is left blank. The dilute
     states possesses a region of stability; the corresponding
     neutral curve is shown as a solid black line in (\emph{a}). 
     Overlaid dashed
     contours indicate lines of constant slope angle.  
     }
\label{fig:morpho existence}
\end{figure}

Finally, we return to the unregularised suspended load model and explore the
$(d,\Fr)$-parameter space more broadly, by computing
the steady states that can exist and their linear stability, when the solid
diameter $d$ is varied over three decades.
We assume here that the non-dimensional settling speed is constant and unity (as
in table~\ref{tab:illustrative params}) throughout, even though its dimensional
counterpart~$\tilde w_s$ varies considerably with the particle size.  This is
reasonable for sufficiently large particles, since~$\tilde w_s \approx \tilde
u_p$, when the particle Reynolds number is high. \citep[This is straightforward
to confirm from typical empirical formulae for $\tilde w_s$, see
e.g.][]{Cheng1997,Soulsby1997}.  Figure~\ref{fig:morpho existence} summarises
the existence of (\emph{a})~dilute and (\emph{b})~concentrated steady flows
across parameter space.
Where steady solutions exist, we contour them according to $\psi_0$.
Elsewhere, we leave the region blank. Overlaid are contours showing lines of
constant slope angle. Care must be taken when interpreting this
plot, since its values depend on the choice of parameters. However, its
qualitative characteristics are robust.
The most striking observation is the separation of the two states,
which are either highly dilute or highly granular. This is clear from
considering the picture in figure~\ref{fig:erodep}.  Solutions exist
predominantly for higher~$d$, where erosion rates are typically smaller and may
therefore balance deposition at higher~$\Fr$. As $\Fr$ increases, keeping~$d$
fixed, eventually~$E_p > D_p$ and states can no longer be steady.  Bounding the
region of existence from above is the unidimensional family of states of type
(\emph{b}) in figure~\ref{fig:erodep}, where the dilute and concentrated
branches coalesce.
Consequently, the dilute and concentrated states respectively possess greater
and lesser solid fractions close to this boundary than they do in the bulk of
parameter space.
On the dilute contour map, figure~\ref{fig:morpho existence}(\emph{a}), we also plot the neutral stability curve (solid
black), below which states are stable to both hydraulic and morphodynamic modes
of disturbance.  It is determined (for our chosen closures) by the asymptotic
growth rate $\lambda_{0,2}$ (which diverges to $+\infty$ at $\Fr = 1$) crossing
zero.
Therefore, we compute it by numerically solving $f_-(\Fr) = 0$, where $f_-$ was given in~\eqref{eq:f
pm}. As indicated by figure~\ref{fig:max growth
vs Fr}(\emph{a}), the upper stability limit lies just below the line $\Fr = 1$
for larger $d$ values.  At smaller $d$, the neutral line dips as it approaches the
existence boundary where solutions are more concentrated. For the parameters
used herein, the region of stability includes only extremely shallow grades
(typically less than~$1^\circ$). However, this is not unexpected, since fluid
drag predominates and purely Ch\'ezy layers turn supercritical (in this case)
when $\phi = \arctan(C_d) \approx 0.6^\circ$.
%
Beneath the region where stable dilute erosive flows exist is a region where
flows are not sufficiently energetic to entrain material, $\theta < \theta_c$,
indicated by a dotted line. Here, only steady flows with zero solid fraction
exist. In the concentrated case, this region is 
only very narrow. Note that for
any $d$, the limiting solutions as $\Fr \to 0$ on this branch are static granular
layers resting at the neutral slope angle $\arctan(\mu_1) \approx 5.7^\circ$
[see~\eqref{eq:pouliquen mu} for the definition of $\mu_1$].
Such states typically have high Shields numbers in excess of the constant
part $\theta_c^*$ of the critical value and consequently any
increase in $\Fr$ from zero leads to entrainment. The remainder of
parameter space in the concentrated case features steady flows at a range of
more severe slope angles, all of which are unstable.

\section{Discussion}
\label{sec:discussion}
This study considered the linear response of spatially uniform steady flows on
constant slopes to small disturbances, in a general class of morphodynamic
shallow-layer models.
Our particular interest was situations where there is significant
entrainment of bed material (assumed to be a saturated mixture of monodisperse
sediment) into the bulk of the flow. 
We therefore focussed on obtaining results for models developed over the
past two decades to describe various highly erosive events such as
violent dam failures, flash floods and volcanic lahars. These models augment
classical shallow-layer formulations used in hydraulic engineering by accounting
for density variations in the flowing mixture, the dynamics of solid transport
and the complex processes of exchange between bed and bulk.
While they do not typically include a separate bed load -- a distinguished layer
that transports sediment along the bed surface (as depicted in
figure~\ref{fig:setup}), we included such a term at various stages to
connect our work with the wider literature on fluvial modelling.
Analysis was performed on a generic set of governing
equations that may be adapted to specific models by specifying (or omitting)
particular closures. 

When entrainment of bed material is significant, the stability picture becomes
substantially modified, compared with past hydraulic analyses
\citep[e.g.][]{Trowbridge1987}, due to the presence of two extra modes of
instability and complicated coupling relationships between hydraulic and
morphodynamic feedbacks. For the suspended load model (negligible bed load), we
derived analytical formulae for the growth rates of normal
mode disturbances in the limits of low and high wavenumber $k$.
%
Most importantly, we observed that the bed evolution equation
gives rise to a zero characteristic wave speed that inevitably intersects with
one of the hydraulic characteristics at $\Fr = 1$, leading to singularities in
the asymptotic (high-$k$) linear growth rates, as observed in
figure~\ref{fig:max growth vs Fr}. Existence of these singularities implies two
important consequences. Firstly, that these models feature a morphodynamic
instability that occurs slightly below unit Froude number.  Secondly, and more
seriously, that the governing equations are ill posed as initial value problems
at $\Fr = 1$, since they permit spatial disturbances to grow arbitrarily rapidly
in the limit $k\to\infty$.

Ill posedness is a critical problem for numerical simulations that must always
be addressed.  However, efforts to solve such models may nonetheless yield
plausible results that match observed properties of real flows.  This is because
the effects of numerical discretisation can make it difficult to identify ill
posedness from isolated results, since the length scales over which severe
disturbances might develop and grow are limited by spatial resolution.  The key
indication is that reference solutions cannot be converged in an ill-posed
system, since finer grid scales only serve to make the discrete system
increasingly sensitive to numerical errors \citep[see][for an example of a
resolution-dependent fingering instability in an ill-posed granular flow
model]{Woodhouse2012}.
Since erosional shallow flow models with solids transport are needed in
critical applications such as hydraulic engineering and natural hazard
assessment, it is vital that their numerical solutions are robust. Consequently,
operational codes that simulate only the basic suspended load model should be
avoided.

In \S\ref{sec:regularisation}, we proved that the inclusion of a simple
turbulence closure (eddy viscosity) suffices to remove ill posedness from the
suspended load model. 
It is therefore tempting to recommend that the equations should always be
regularised with at
least a small amount of eddy viscosity.
However, even a small amount of diffusion
changes the fundamental structure of the model equations and may make them more
difficult to time step in a numerical code. 
Nevertheless, at least one study (within our general framework) includes this
term~\citep{Simpson2006}. Moreover, we might anticipate that other turbulence
closures, or analogous diffusive terms such as those employed in recent shallow
granular flow models~\citep{GrayEdwards2014}, similarly avoid ill posedness by
damping growth in the short-wave limit. As shown in~\S\ref{sec:effect of
regularisation}, the morphodynamic instability near $\Fr = 1$ persists when the
model is regularised by eddy viscosity and its onset is unaffected if the
regularising term is small.  However, figure~\ref{fig:max growth vs Fr visc}
demonstrated that the magnitude of the eddy viscosity has a significant impact
on the severity of this instability. Therefore, selection and calibration of
a suitable diffusive closure is far from arbitrary, since it could dictate whether
instabilities are seen over the finite lifetime of a simulated geophysical flow.
A full investigation of such terms would require careful
comparisons with experimental or observed flows.

The removal of ill posedness, through the introduction of eddy viscosity or bed
load flux (as in~\S\ref{sec:effect of bed load} and discussed below), does not
imply removal of the associated morphodynamic instability
that arises near $\Fr = 1$. We have not speculated much about the
physics of this instability in the main body of the paper. It may be that it is
a purely artificial phenomenon, whose relevance disappears when models are
properly calibrated and include all physically important processes.  However, in
the extended analyses of our illustrative closures, including the eddy viscosity
and bed load terms, we were not able to rule out the
destabilising influence of the $\Fr = 1$ singularity.
Hence, both the severity of its growth and its presence
at modest Froude numbers ($\Fr \gtrsim 1$), 
make it a feature that should be carefully considered when employing these
models.
Since the
morphodynamic instability exists essentially due to a resonance between the free
surfaces of the flow and the bed at short wavelengths, its early development
should feature rapid growth of fine scale structure in these fields.  Indeed,
the growth of mode~IV is typically dominant -- its components in the asymptotic
limit, given by the final vector in~\eqref{eq:O(1) eigvecs}, couple high
frequency oscillations in~$h$, $u$ and~$b$.  To precisely confirm the onset of
this instability in a concentrated geophysical flow or a relevant experimental
set-up would be challenging.  Moreover, while similar resonances have been
studied in morphodynamic potential flow models, the resulting instabilities were
found to disappear when more detailed physical models were
employed~\citep{Coleman2000,Colombini2005}.  
However, regardless of these uncertainties, a detailed understanding of the
morphodynamic instability is required in order to make properly informed
modelling decisions and may be used to guide future model development.

An important next step would be to investigate how the instability develops
beyond the linear regime.
This could be assessed by conducting a careful nonlinear analysis in the vein of
\cite{Needham1984}, or via carefully resolved numerical simulations of a
suitably regularised system.  As observed in~\S\ref{sec:effect of eps}, the
associated suspended load dynamics acts to suppress the growth rate of mode~I,
which is responsible for roll wave instability in the hydraulic limit.  On this
basis, we speculate that the morphodynamic instability may not ultimately
cause the flow to roll up into large free-surface waves.
%
Instead, it seems more closely related
to an upstream-propagating bedform instability
discovered by~\cite{Balmforth2012}
in 
%
a simplified model, where $\Gamma$ is assumed to be negligible in all but the
bed equation and $Q = 0$.
This formulation 
also suffers a
singularity at $\Fr = 1$, unless turbulent momentum diffusion is included. 
The eddy
viscosities used to regularise their system were $\nu \sim 10^{-2}$ to
$10^{-1}$ -- large enough to subdue any dramatic short-wave growth arising from
the singularity.  However, when $\nu = 10^{-2}$, our formulation is nonetheless
morphodynamically unstable for all $\Fr \gtrsim 1.05$ (see figure~\ref{fig:max
growth vs Fr visc}).  Moreover, an illustrative numerical calculation (not
shown) indicates that it is indeed the mode associated with the bedform that
turns unstable (consistent with the role of mode~IV elsewhere), with slow
upstream-directed phase speed (\eg $c = -0.021$ for the most unstable mode, when
$\Fr = 1.2$).  It seems reasonable to expect that lower effective turbulent
viscosities will be present in at least some natural morphodynamic systems. (See
the discussion closing~\S\ref{sec:effect of regularisation} for an estimate of
the range of~$\nu$.) Whether or not this leads to the more severe instabilities
predicted by some of our results remains to be established.

Bed load is an important physical process whose inclusion, via a flux term $Q$
in the basal dynamics equation~\eqref{eq:mass transfer}, modifies the
characteristic wave speeds of the governing equations. It therefore plays a key
role in determining whether the model is strictly hyperbolic (and consequently
well posed) or not. This is already well appreciated in models of river
morphodynamics, where bed load fluxes are frequently employed and the case for
one or more sediment transport layers near the bed surface is experimentally and
observationally clear.  Consequently, a number of recent studies have
investigated conditions for well posedness in these
settings~\citep{Cordier2011,Stecca2014,Chavarrias2018,Chavarrias2019}.
Conversely, models of shallow highly concentrated suspensions rarely include a
bed load, since these flows are typically feature an energetic and well-mixed
bulk.  However, a better approach may be to consider flows on a continuum, from
a dilute bed load regime to highly concentrated suspensions. While increasing
Shields number causes more grains to be carried into suspension, it seems
unreasonable to conclude that $Q$ ultimately shuts off and the bed characteristic
becomes zero.
Therefore, it may always be prudent to include a bed flux term, in order to
avoid potentially artificial resonance between the hydraulic and morphodynamic
modes. 
Such models could easily be checked against the criterion derived
in~\eqref{eq:hyperbolic ineq}
to ensure that they are well posed.

An investigation of the effects of bed load with example model closures
in~\S\ref{sec:effect of bed load} demonstrated that its effect on growth rates
is similar to that of eddy viscosity -- it mollifies the acute growth
rates around the $\Fr = 1$ singularity and increases the critical $\Fr$ for
instability.
%
Our results in figure~\ref{fig:max growth vs Fr bedload} suggest that both the
severity and dominant mode of instability are determined through competition
between the morphodynamics of the suspended and bed loads.  Indeed, since
morphodynamic instabilities are not present near $\Fr = 1$ in pure bed load
models~\citep{Lanzoni2006}, the effect of mass transfer with the suspended load
appears to be destabilising. The predicted instabilities in the $1 \lesssim \Fr
\lesssim 2$ region migrate slowly upstream. This is in qualitative agreement
with fluvial models that do not employ the shallow flow approximation~\citep[see amongst
others,][]{Engelund1970,Colombini2004,Colombini2008,Seminara2010}.  These models
capture a richer variety of pattern-forming instabilities than appear to be
accessible to shallow formulations, such as the formation of dunes for $\Fr < 1$
and various other features~\citep{Richards1980,Seminara2010,Colombini2011}.
Nevertheless, it is interesting that the combined (bed and suspended load)
formulation exhibits some morphodynamic instabilities. 

Finally, in \S\ref{sec:existence} we demonstrated, that steady morphodynamic
layers (when they exist) bifurcate into two coexistent states: dilute stable
layers and concentrated unstable layers.  This is a basic physical idea that
lies apart from issues of model consistency and is largely independent of the
model closures.  In essence, the solutions arise due to the effects of hindered
settling, which render the deposition rate non-monotonic with respect to the
bulk solid fraction. This means that there are two possible sediment
concentrations where erosion exactly balances deposition. Above a certain
threshold of $\Fr$, both states cease to exist, since erosion everywhere exceeds
the maximal rate of deposition. For the most part, simple physical arguments
suffice to explain the stability of the two branches (see~\S\ref{sec:existence}
and~\ref{sec:global morpho}), as we were able to confirm via careful analysis of
the linear growth rates in~\S\ref{sec:global morpho}. This general picture
appears to accord with observations of natural flows. Both natural and
laboratory debris flows often propagate as an unsteady surge-like front followed
by a shallow stable layer of weaker sediment concentration, with this
configuration repeating during the
flow~\citep{Davies1992,Zanuttigh2007,Doyle2010}.  For instance, flows of
volcanic debris (lahars) typically propagate as alternating debris-rich pulses
and relatively shallower and less concentrated ($\sim20\%$ by volume solids
concentration) layers~\citep{Pierson2005,Doyle2010}.
Previous studies have used linear stability analysis to explore the link between
flow instabilities such as roll waves and the development of pulses in debris
flows~\citep[e.g.][]{Zanuttigh2007}. Further study including fully nonlinear
analysis of morphodynamic shallow-layer models, is needed in order to properly
link the mechanisms in this paper with observations of natural flows and
provides an interesting opportunity for future research.

\acknowledgements{
We thank C.\ G.\ Johnson, University of Manchester, for useful discussions
concerning shallow-layer models and analysis, and L.\ T.\ Jenkins for comments
on the manuscript.
The main results of this paper were obtained as part of the Newton Fund
grant `Quantitative Lahar Impact and Loss Assessment under Changing Land
Use and Climate Scenarios': NE/S00274X/1. Initial investigations were
conducted during the ‘Strengthening Resilience in Volcanic Areas’
(STREVA) project, funded by the Natural Environment Research Council
(NERC) and the Economic and Social Research Council (ESRC): NE/J020052/1.
MJW acknowledges funding from the NERC award `VolcTools -- enhancing ease of use
and uptake of tools to improve prediction and preparedness of volcanic hazards':
NE/R003890/1;
AJH acknowledges an APEX fellowship from the Royal Society, UK: APX/R1/180148;
and
JCP acknowledges support from a University of Bristol Research Fellowship.
}

~\\
The authors report no conflict of interest.

\bibliographystyle{jfm}
\bibliography{LaharBib}

\end{document}